\documentclass[11pt]{article}

\newcommand{\mayaprl}{\raisebox{-.3ex}{$\enskip\stackrel{<}
{\scriptstyle\sim}\enskip$}}

\usepackage[dvips]{graphicx}
\usepackage{epsfig}

\begin{document}
\begin{center}
{\LARGE A Systematic Expansion Method
in Granular Hydrodynamics}\\[1.cm]
J. Wakou\footnote{
The present address is:
Miyakonojo National College of Technology,
473-1 Yoshio-cho, Miyakonojo, Miyazaki 885-8567, Japan; E-mail: wakou@miyakonojo-nct.ac.jp.
}
\\
{\small\it
Instituut voor Theoretische Fysica, Universiteit Utrecht,
P.O.Box 80.195, 3508 TD Utrecht, The Netherlands}\\
\end{center}

\section*{Abstract}
We propose a systematic expansion method which is applied to freely evolving
granular fluids contained in sufficiently small systems.
Restricting ourselves to small systems, we show that there exists
a small parameter which characterizes
a typical size of density and temperature inhomogeneities. A solution
of the hydrodynamic equations for a fluid of inelastic hard spheres
is expanded in the small parameter. It is shown that our method
quantitatively describes the asymptotic state of the system, such as flow profiles,
density and temperature inhomogeneities,
and the decay law of the global temperature and the energy per particle.

\section*{PACS numbers}
45.70.Mg, 45.70.Qj, 47.85.Dh, 02.30.Mv.

\section*{Key words}
Granular fluid; instabilities; pattern formation;
hydrodynamic equations; asymptotic state; perturbation expansion.

\section{Introduction}
Granular matter shows fluid like behavior under certain
conditions~\cite{campbell,jaeger-nagel}, such as Couette flow,
convection flow under vibration,
and flow down an inclined chute.
A freely cooling rapid granular flow is an idealized
limiting case of granular fluids in the absence of gravity and
without any energy input.
By analyzing their molecular dynamics (MD) simulations of a system
of smooth inelastic hard discs,
Goldhirsch and Zanetti~\cite{goldhirsch,goldhirsch2} discovered
that the freely cooling granular
fluids exhibit interesting instabilities. When the system is
prepared in a spatially homogeneous state, it slowly develops
patterns in the flow field (vortices) and in the density field
(clusters).

A linear stability analysis~\cite{goldhirsch,goldhirsch2,mcnamara,deltour-barrat,brey1,vn-er-pre}
of hydrodynamic equations for inelastic hard spheres (IHS) has
revealed that the initial spatially homogeneous cooling state
is unstable to the formation of vortices and clusters,
and provided
a good description of the initial growth of unstable hydrodynamic modes.
At large times, however, these unstable hydrodynamic modes
grow so large that nonlinear hydrodynamic effects, such as
convection and viscous heating, may dominate the evolution of
the system.
Nonlinear effects in freely cooling granular fluids have been discussed
by several authors~\cite{goldhirsch,goldhirsch2,brey,ben-naim,soto,santafe}.
In Refs.~\cite{goldhirsch,goldhirsch2}, the
importance of viscous heating effects has been pointed out and
the threshold of the clustering instability has been given
in terms of the coefficient of restitution and a typical length scale of a fluctuation.
Nonlinear coupling between hydrodynamic modes through
viscous heating has been studied~\cite{brey}
by means of numerical analysis of a hydrodynamic
model and direct Monte Carlo simulation (DMCS) of the Boltzmann equation.
Recently, strong numerical evidence has been shown~\cite{ben-naim} that
a one-dimensional freely cooling granular
gas asymptotically behaves as a sticky gas, and that the inviscid Burgers
equation is an appropriate continuum theory.
These three works are concerned with unstable growth regime in large systems, which is
the physically most interesting problem of hydrodynamics for the freely cooling
granular fluids. This regime is characterized by intrinsic patterns
in the flow field and in the density field where the typical correlation length
of these patterns are small compared to the system size.
Unfortunately in this regime only analytical large time results have been
obtained from a mode coupling theory~\cite{brito-ernst,brito-ernst2,chen}.

Recently some progress in the analysis of {\it nonlinear}
hydrodynamic equations
has been made for freely cooling granular fluids contained in systems
that are {\it sufficiently small}, such that
the clustering instability is suppressed~\cite{soto,santafe}.
A linear stability
analysis~\cite{goldhirsch,goldhirsch2,mcnamara,deltour-barrat,brey1,vn-er-pre}
predicts that if the smallest wave number of a system,
$k_0=2\pi/L$, is smaller than a critical value
$k_{\perp}^{*}$,
the homogeneous cooling state becomes unstable and
the flow field develops
a shear flow pattern.
Several groups have observed
by MD simulations~\cite{goldhirsch,goldhirsch2,deltour-barrat,obne,soto,santafe,mcnamara-young}
and by DMCS~\cite{brey} of IHS contained in two-dimensional systems with periodic boundaries
that there is a regime of $k_0$ for a given coefficient of
restitution where the system develops
shearing patterns, but no strong density inhomogeneities.
In this regime the flow field
eventually reaches a stationary shearing pattern that has
the following properties:
(i) a stationary shear flow profile with a period equal to
the system size in the direction perpendicular
to the direction of the flow~\cite{goldhirsch,goldhirsch2,deltour-barrat,obne,soto,santafe,mcnamara-young},
(ii) exponential decay of the energy per particle as a function of $\tau$,
the average number of collisions suffered per particle in time $t$~\cite{goldhirsch,goldhirsch2,santafe,b-o},
(iii) $t^{-2}$-decay of the energy per particle~\cite{santafe,b-o},
(iv) inhomogeneities in density~\cite{goldhirsch,goldhirsch2,brey,soto,santafe,mcnamara-young} and temperature~\cite{santafe}
with a period that is half
the period of the shear flow profile; 
the density concentrates into regions of the peaks of the flow profile, 
and
(v) saturation of the density profile~\cite{brey}.
The properties (i)-(iii) have also been observed by MD simulations in five and six dimensions~\cite{tri-bar}.
In Ref.~\cite{soto}, the evolution of shear flow patterns
in systems with $k_0\sim k_{\perp}^{*}$ ($k_0< k_{\perp}^{*}$)
has been studied.
In such systems, unstable shear modes grow so slowly that
the remaining hydrodynamic modes are enslaved by
the shear modes, and the amplitudes of the unstable shear modes
remain small. On the basis of this consideration, they have derived
amplitude equations for the unstable shear modes
and explained the above properties in a quantitative manner.
For similar small systems, in Ref.~\cite{santafe}
a Landau-Ginzburg-type equation for a nonconserved
order parameter, the shear rate tensor, has been derived
under an approximation in which inhomogeneities in
density and temperature are neglected.
It has been shown that the Landau-Ginzburg-type
equation describes saturation of unstable
shear modes due to viscous heating effects, and describes
the properties of shear flow profile (i)-(iii) at large times
in a quantitative manner
in spite of drastic simplification achieved by the approximation.

This paper is also concerned with the small system regime;
we propose a systematic expansion method to obtain solutions
which correspond to the asymptotic state of sufficiently small systems.
The method is applied to the stationary shear flow profile with
the properties (i)-(v), which appears in sufficiently small systems
with periodic boundaries. Although the asymptotic state
of the small systems with periodic boudaries is not a physical state,
we concentrate in this paper only on this case, which has been studied
well so far by the computer simulations~\cite{goldhirsch,goldhirsch2,deltour-barrat,obne,soto,santafe,mcnamara-young} 
and by the theories~\cite{soto,santafe}.
The method is based on an expansion in the Prandtl number $P_r$
of a solution of the hydrodynamic equations for IHS.
At large times in small systems, viscous heating, which is proportional to the kinematic
viscosity $\nu$, due to growing flow field patterns
becomes dominating effects that induce inhomogeneities in temperature~\cite{goldhirsch,goldhirsch2,brey,soto,santafe}.
In sufficiently small systems, the inhomogeneities in temperature
is largely reduced by thermal conduction, which is proportional to
the thermal conductivity $\kappa$.
In these small systems, a typical size of temperature inhomogeneities
can be determined by these two competing effects; hence it is proportional to
the Prandtl number $P_r\propto\nu/\kappa$.
We will show that by expanding the solutions in the Prandtl number,
the inhomogeneities in density and temperature field in the small system
regime can be described;
corrections to shear flow profile, and the energy
and temperature decay laws caused by these density and
temperature inhomogeneities are obtained in a systematic manner.
This method has the approximation used in Ref. \cite{santafe}
that density and temperature
inhomogeneities are neglected as its zeroth approximation.
Our method can describe systems
with finite amplitudes of unstable shear modes,
to which the method in Ref.~\cite{soto} can not be applied.
The idea of expanding solutions of hydrodynamic equations in a small Prandtl number
was originally developed for classical fluids by Busse~\cite{busse}
in order to describe an oscillatory instability of convection rolls
that appear in a fluid layer heated from below.

This paper is organized as follows. In Sec.2, we introduce the hydrodynamic
equations for IHS and show that we can eliminate the global temperature,
defined as the spatial average of the local temperature,
from the hydrodynamic equations by introducing rescaled hydrodynamic variables.
Then, it is shown that in the small system regime some of the properties (i)-(v)
can be associated with the existence of a stationary solution of
the hydrodynamic equations for the rescaled variables.
In Sec.3, the method of systematic expansion in the Prandtl number is presented.
We first discuss the basic idea of the expansion scheme,
then show how the expansion scheme works up to the second
approximation.
A quantitative description of the properties (i)-(v) is given in the
first approximation. It presents qualitatively new predictions
of the behavior of $d$-dimensional IHS fluids.
The range of validity of our method is discussed
in the end of Sec.3. We end with some conclusions in Sec.4.

\section{Hydrodynamics of IHS}
\subsection{Hydrodynamic equations for IHS}
We consider a system of $N$ inelastic hard spheres,
contained in a $d$-dimensional cubic system with
sides $L$ and volume $V=L^d$.
The boundary conditions are periodic boundary conditions.
At an initial time, $t=0$, the system is prepared in a homogeneous
state with temperature $T_0$ and vanishing flow field.
We assume that weekly inelastic hard sphere systems
can be described by the hydrodynamic equations
supplemented
by a term $\Gamma$ that expresses the rate of collisional energy loss~\cite{jenkins-richman,jenkins-richman2},
\begin{eqnarray}
\partial_t n +{\bf u}\cdot\nabla n &=& -n\,\nabla\cdot {\bf u} , \label{density}
\\
\partial_t {\bf u} +{\bf u}\cdot\nabla {\bf u} &=& -\frac{1}{mn}\nabla p +\frac{2}{m
n}\nabla\cdot\left(\eta \,\mbox{\sf D}\right) +\frac{1}{m n}\nabla
\left(\zeta\nabla\cdot{\bf u}\right) , \label{velocity}
\\
\partial_t T +{\bf u}\cdot\nabla T &=& -\frac{2p}{dn}\nabla\cdot {\bf u}
+\frac{2}{dn}\nabla\cdot(\kappa\nabla T)
\nonumber\\
&+&\frac{4\eta}{dn}\left(\mbox{\sf D}:\mbox{\sf D}\right)
+\frac{2\zeta}{dn}(\nabla\cdot{\bf u})^2 -\Gamma .
\label{temperature}
\end{eqnarray}
In this paper the inelasticity is always assumed to be small.
The shear rate $\mbox{\sf D}$  is the symmetrized dyadic,
$\mbox{\sf D}_{\alpha\beta}=(\nabla_{\alpha}u_{\beta}
+\nabla_{\beta}u_{\alpha}
-\frac{2}{d}\delta_{\alpha\beta}\nabla\cdot{\bf u})/2$,
and $A:B=\sum_{\alpha=1}^{d}\sum_{\beta=1}^{d}A_{\alpha\beta}B_{\beta\alpha}$.
The transport coefficients, the shear viscosity $\eta$, the bulk
viscosity $\zeta$, and the heat conductivity $\kappa$, depend on
the local density and temperature.

We observe that the total energy per particle consists of
a convective and internal energy
\begin{eqnarray}
E=E_{conv}+E_{int}=\frac{1}{N}\int d{\bf r}
\left[\frac{1}{2}m n u^2+\frac{d}{2}n T\right]
,
\end{eqnarray}
with rate of change, derived from Eqs. (\ref{density})-(\ref{temperature})
\begin{eqnarray}
\frac{d E}{d t}&=&-\frac{1}{N}\int d {\bf r}\frac{d}{2}n\Gamma
,
\label{dedt}
\\
\frac{d E_{int}}{d t}&=&\frac{1}{N}\int d {\bf r}
\left[-p\nabla\cdot{\bf u}+2\eta\mbox{\sf D}:\mbox{\sf D}
+\zeta(\nabla\cdot{\bf u})^2-\frac{d}{2}n\Gamma\right]
.
\label{deconvdt}
\end{eqnarray}
The terms on the right hand side of the last equation
represent the work done by the flow, the gain of $E_{int}$
due to viscous heating and the loss of $E_{int}$ due to collisional dissipation.

On the basis of kinetic theory one can derive that the rate of
collisional energy loss, $\Gamma = 2 \gamma_0\omega T$, is
proportional to the collision frequency $\omega(n,T)$ multiplied by the
fraction of energy $\epsilon T$ lost per collision \cite{jenkins-richman,jenkins-richman2},
where $\gamma_0=\epsilon/2d=(1-\alpha^2)/2d$ is the inelasticity parameter
and $\alpha$ is the coefficient of restitution.

\subsection{Elimination of the global temperature}
In this section, we will introduce
a rescaling of the hydrodynamic variables
by means of the global granular temperature defined as
\begin{eqnarray}
\bar{T}(t)=\frac{2}{d}E_{int}(t)
=\frac{1}{N}\int d{\bf r}\,n({\bf r},t)\,T({\bf r},t)
,
\label{gtemp}
\end{eqnarray}
or equivalently the mean random (thermal) velocity
$\bar{v}_0=\sqrt{2\bar{T}/m}$. Similarly we will use
the global density $\bar{n}=N/V$.

The freely cooling hard sphere fluid has no intrinsic energy
scale: there is no energy scale in parameters which characterize
interactions between particles and interactions between a particle
and a boundary. Hence, if we rescale time by
$t\mapsto a t$ at fixed parameters that characterize interactions
between particles and interactions between a particle and a boundary,
then velocities of all particles, energy, and temperature are
rescaled as $v\mapsto a^{-1}v$, $E\mapsto a^{-2}E$, and $T\mapsto a^{-2}T$,
respectively,
but the trajectories of particles are not affected by this rescaling.
%
%
Therefore, we can always keep one of energy scales in the system
constant by rescaling time in an appropriate way.
This fact has been utilized by Soto et al.~\cite{soto}
to enhance accuracy of their MD simulations by keeping the total energy constant.
In this article, keeping the global temperature constant
is suitable for our purpose because
the collision frequency and the transport coefficients
in the hydrodynamic equations for IHS are given as functions of
temperature and density.
The rescaling of the hydrodynamic variables given below has been used
in a linear stability analysis~\cite{goldhirsch,goldhirsch2,mcnamara,deltour-barrat,brey1,vn-er-pre}.

We first introduce the dimensionless time and the scaled
hydrodynamic variables
\begin{eqnarray}
d\tau &=&\bar{\omega}\,d t, \hspace{0.5cm}
\hspace{0.5cm} \tilde{n} =\frac{n}{\bar{n}}, \hspace{0.5cm}
\tilde{{\bf u}}=\frac{{\bf u}}{\bar{v}_0}, \hspace{0.5cm}
\tilde{T}=\frac{T}{\bar{T}} ,
\label{dimless}
\end{eqnarray}
where $\bar{\omega}=\omega(\bar{n},\bar{T})$ is the
collision frequency taken at the global temperature $\bar{T}(t)$ and
the global density $\bar{n}$,
and $\tau$ is the average number of collisions, suffered
by a particle in time $t$.
As a result of the rescaling, the scaled global temperature
$(1/N)\int d {\bf r} \tilde{n}\tilde{T}$ always remains constant.
The collision frequency $\omega$ and the transport coefficients $\{\eta,\zeta,\kappa\}$
in the hydrodynamic equations will be calculated from
the Enskog theory for hard spheres, in principle for
inelastic ones~\cite{goldstein-sapiro,brey-dufty-kim-santos,garso-dufty}.
Moreover,
$\bar{\eta}=\eta(\bar{n},\bar{T})$,
$\bar{\zeta}=\zeta(\bar{n},\bar{T})$,
$\bar{\kappa}=\kappa(\bar{n},\bar{T})$
denote the corresponding Enskog transport coefficients
taken at $\bar{T}(t)$ and $\bar{n}$.
Note that in the Enskog theory for hard spheres,
$\bar{\omega},\bar{\eta},\bar{\zeta},\bar{\kappa}\propto\sqrt{\bar{T}}$.
Then, we get a closed set of equations for these
scaled variables,
\begin{eqnarray}
\partial_\tau \tilde{n}+\tilde{{\bf u}}\cdot l_0\nabla \tilde{n}
&=&
-\tilde{n}\,l_0\nabla\cdot
\tilde{{\bf u}}
,
\label{sdensity}
\\
\partial_\tau \tilde{{\bf u}}+\tilde{{\bf u}}\cdot l_0\nabla \tilde{{\bf u}}
&=&
-\frac{1}{2\tilde{n}}l_0\nabla \tilde{p}
+D_{\perp}
\frac{2}{\tilde{n}}\nabla\cdot\left(\eta^{*}(\tilde{n})
\sqrt{\tilde{T}} \,\tilde{\mbox{\sf D}}\right)
+\frac{\bar{\zeta}}{\bar{\rho}\bar{\omega}}
\frac{1}{\tilde{n}}\nabla
\left(\zeta^{*}(\tilde{n})\sqrt{\tilde{T}}\nabla\cdot\tilde{{\bf u}}\right)
\nonumber\\
&-&
\frac{1}{2}\left(\partial_\tau \ln
\bar{T}\right)\tilde{{\bf u}}
,
\label{svelocity}
\\
\partial_\tau \tilde{T} +\tilde{{\bf u}}\cdot l_0\nabla \tilde{T}
&=&
-\frac{2}{d}\frac{\tilde{p}}{\tilde{n}}l_0\nabla\cdot \tilde{{\bf u}}
+\frac{\bar{c}_p}{c_v} D_T \frac{1}{\tilde{n}}
\nabla\cdot\left(\kappa^{*}(\tilde{n})\sqrt{\tilde{T}}\nabla
\tilde{T}\right)
\nonumber\\
&+& \frac{8}{d}D_{\perp}
\frac{\eta^{*}(\tilde{n})}{\tilde{n}}\sqrt{\tilde{T}}
\left(\tilde{\mbox{\sf D}}:\tilde{\mbox{\sf D}}\right)
+\frac{4}{d}\frac{\bar{\zeta}}{\bar{\rho}\bar{\omega}}
\frac{\zeta^{*}(\tilde{n})}{\tilde{n}}\sqrt{\tilde{T}}
(\nabla\cdot\tilde{{\bf u}})^2
\nonumber\\
&-&
2\gamma_0\omega^{*}(\tilde{n})\tilde{T}^{3/2}
-\left(\partial_{\tau} \ln \bar{T}\right)\tilde{T}
,
\label{stemperature}
\end{eqnarray}
where we have introduced the following notation:
\begin{eqnarray}
\frac{\eta}{\bar{\eta}}&=&\eta^{*}(\tilde{n})\sqrt{\tilde{T}},\hspace{0.5cm}
\frac{\zeta}{\bar{\zeta}}=\zeta^{*}(\tilde{n})\sqrt{\tilde{T}},\hspace{0.5cm}
\frac{\kappa}{\bar{\kappa}}=\kappa^{*}(\tilde{n})\sqrt{\tilde{T}},\hspace{0.5cm}
\\
\frac{\omega}{\bar{\omega}}&=&\omega^{*}(\tilde{n})\sqrt{\tilde{T}},\hspace{0.5cm}
\frac{p}{\bar{n}\bar{T}}=\tilde{n}\tilde{T}\,p^{*}(\tilde{n})=\tilde{p}(\tilde{n},\tilde{T})
.
\end{eqnarray}
The functions $f^{*}(\tilde{n})$ depend on $\tilde{n}$ and
the coefficient of restitution $\alpha$,
and $\eta^{*}(1)=\zeta^{*}(1)=\kappa^{*}(1)=\omega^{*}(1)=1$.
The constants $D_\perp=\bar{\nu}/\bar{\omega}$ and
$D_T=\bar{\kappa}/\bar{c}_p\bar{\rho}\bar{\omega}$ are the rescaled shear diffusivity
and the heat diffusivity, respectively, and
$c_v=d/2m$ and $\bar{c}_p=c_p(\bar{n})$ are
the specific heat per unit mass at constant volume and pressure.
The mean free path $l_0$ is defined by $l_0=\bar{v}_0/\bar{\omega}_0$.
It is important to note that the constants $D_{\perp}, D_{T},\bar{\zeta}/\bar{\rho}\bar{\omega}$ and
$l_0$ are independent of $\bar{T}$.
The macroscopic equation for the global temperature $\bar{T}$
can be obtained from $d E_{int}/dt$ in Eq.~(\ref{deconvdt}) and subsequent rescaling,
and yields,
\begin{eqnarray}
\partial_\tau \ln \bar{T}
&=&
\frac{1}{V}\int d{\bf r}\biggl[
-\frac{2}{d}\tilde{p}l_0\nabla\cdot \tilde{{\bf u}}
+\frac{8}{d}D_{\perp}\eta^{*}(\tilde{n})\sqrt{\tilde{T}}
\left(\tilde{\mbox{\sf D}}:\tilde{\mbox{\sf D}}\right)
\nonumber\\
&&
+\frac{4}{d}\frac{\bar{\zeta}}{\bar{\rho}\bar{\omega}}
\zeta^{*}(\tilde{n})\sqrt{\tilde{T}}
\left(\nabla\cdot\tilde{{\bf u}}\right)^2 -
2\gamma_0\omega^{*}(\tilde{n})\tilde{n}\tilde{T}^{3/2} \biggr]
.
\label{eqgtemp}
\end{eqnarray}
Hence, the time derivative of $\ln\bar{T}$ is a functional of
the scaled hydrodynamic variables $\tilde{n},\tilde{{\bf u}}$ and $\tilde{T}$,
and the global temperature can be completely eliminated from
the equations for these scaled hydrodynamic variables.

\subsection{Asymptotic state}
Equations (\ref{sdensity})-(\ref{stemperature}) for the freely evolving
IHS-fluid are supposedly describing the growing patterns in the
density and flow fields starting from a spatially homogeneous initial state,
and the basic objective is to describe their asymptotic evolution for
small inelasticity, where the time scale of collisional cooling is well separated
for the free time between collisions.
Comparison of scaled field Eqs. (\ref{sdensity})-(\ref{stemperature}) with
Eq. (\ref{eqgtemp}) for the global temperature
suggests that there might be two totally different scenarios leading to an asymptotic
state. One small system scenario, where the fields equations describe a fast process
with a typical scale $\tau_{\perp}\simeq L^2/D_{\perp}$. For time $\tau$ large compared to
$\tau_{\perp}$ the scaled hydrodynamic variables reach a stationary solution and
the rate of change of the global temperature, i.e. the right hand side of
Eq. (\ref{eqgtemp}), approaches a nonvanishing constant.
In the large system scenario, the field equations contain arbitrary slow hydrodynamic
modes. The system, in which the global temperature initially decays like
$\exp(-2\gamma_0\tau)$ according to Haff's law~\cite{haff},
gradually reaches an asymptotic state with inhomogeneities in the flow field,
where the viscous heating almost entirely compensates the collisional cooling,
and where the right hand side of Eq.~(\ref{eqgtemp}) becomes very small.
Some consequences of the second scenario have been analyzed in the mode coupling
calculation of Ref.~\cite{brito-ernst,brito-ernst2}.

Here we concentrate on a sufficiently large time scale, $\tau\gg\tau_{\perp}$,
in the small system scenario.
There are some important observations that can be deduced from
the assumption of the {\it existence of a stable stationary
solution} of the scaled hydrodynamic equations.
The results under the approximation in which inhomogeneities in density and
temperature are neglected
have already been reported in Ref.~\cite{santafe}.
The assumption of the existence of a stable stationary solution implies that the right hand side of Eq.~(\ref{eqgtemp}) is constant,
$(\partial_{\tau}\ln\bar{T})\equiv -2\gamma_a <0$. This leads the following important consequences: First, for large $\tau$
the global temperature $\bar{T}$ in Eq.~(\ref{eqgtemp})
decays exponentially in terms of $\tau$,
i.e.
\begin{eqnarray}
\bar{T}\propto e^{-2\gamma_a \tau}.
\label{temptau}
\end{eqnarray}
Second, the exponential decay (\ref{temptau})
of the global temperature implies its $t^{-2}$-decay in the real time $t$.
According to the definition of $\tau$
given in Eq.~(\ref{dimless}), the evolution of the global temperature
$\bar{T}$ in terms of $t$ becomes,
\begin{eqnarray}
\partial_t \bar{T}
&=& \frac{d \tau}{d t}\,\partial_{\tau}\bar{T}
=\bar{\omega}\bar{T}\,(\partial_{\tau}\ln \bar{T})
=\frac{\omega_0}{\sqrt{T_0}}\,(\partial_{\tau}\ln
\bar{T})\,\bar{T}^{3/2} , \label{tempt}
\end{eqnarray}
where $\omega_0(\bar{n},T_0)$ is the collision frequency at
the initial time and $T_0$ is the initial temperature.
Here we have used $\bar{\omega}\propto\sqrt{\bar{T}}$.
At sufficiently large times $(\partial_{\tau}\ln \bar{T})$ can be
replaced by its stationary value $-2\gamma_a$, i.e.
\begin{eqnarray}
\partial_t \bar{T}
&=&-\frac{\omega_0}{\sqrt{T_0}}\,2\gamma_a\,\bar{T}^{3/2} .
\end{eqnarray}
This equation implies that $\bar{T}$ decays as $t^{-2}$ at large
times, which has been indeed observed in simulations. More precisely
\begin{eqnarray}
\bar{T}(t) = \frac{T_0}{\gamma_a^2 \omega_0^2 (t-t_e)^2}\simeq
\frac{T_0}{\gamma_a^2\omega_0^2} t^{-2},
\end{eqnarray}
where $t_e$ is an integration constant.
It should be noted that the prefactor of $t^{-2}$ is a constant which is
independent of transient behavior of the system
and of the initial temperature.
We compare this decay with the initial decay in the homogeneous cooling
state (Haff's law)~\cite{haff}
\begin{eqnarray}
\bar{T}(t)=T_0 \,e^{-2\gamma_0\tau}
=
\frac{T_0}{(1+\gamma_0\omega_0 t)^2}\simeq
\frac{T_0}{\gamma_0^2\omega_0^2} t^{-2},
\end{eqnarray}
where the decay rate in $\tau$ is larger because $\gamma_0>\gamma_a$~\cite{santafe}.
Third, because the collision frequency $\bar{\omega}$ is proportional to
the square root of the global temperature
$\omega\propto\sqrt{\bar{T}}$,
the relation (\ref{dimless}) between $\tau$ and $t$
in the asymptotic state is given by
\begin{eqnarray}
d t  &\propto& \frac{1}{\sqrt{\bar{T}}}d \tau
\propto e^{\gamma_a\tau} d \tau .
\end{eqnarray}
Finally, it can be shown that a stationary solution of the scaled hydrodynamic variables
implies that the energy per particle is proportional to the global temperature:
\begin{eqnarray}
E
&=&
\frac{1}{N} \int d{\bf r}\left[\frac{m}{2}n{\bf u}^2+\frac{d}{2}n T\right]
=
\bar{T}\left[\frac{1}{V}\int d{\bf r}\,\tilde{n}\tilde{{\bf u}}^2+\frac{d}{2}\right]
.
\label{prop}
\end{eqnarray}

These results seem to be consistent with the simulation results (i)-(v) in Sec.1.
Hence, we shall assume that for sufficiently small systems Eqs.
(\ref{sdensity})-(\ref{stemperature}) have a stable stationary
solution, and concentrate on how to obtain the stationary
solution and the exponent $\gamma_a$ by means of an systematic approximation method.

It is important to note that the above argument is valid
also for systems surrounded by inelastic walls as well as
for systems with periodic boundaries; both
systems have no intrinsic energy scale.

\section{Systematic expansion method}
\subsection{Basic idea}
In this section we present the basic idea of a systematic expansion method to
solve Eqs. (\ref{sdensity})-(\ref{stemperature}) for the stationary state, i.e.
with the vanishing time derivatives.
In this argument we shall assume that viscous heating plays an essential role
in inducing inhomogeneities in the density and temperature fields
on the basis of
the previous studies~\cite{goldhirsch,goldhirsch2,brey,soto,santafe}
which have shown importance of viscous heating effects in the formation
of inhomogeneities.
It will be shown later that the result obtained by
the systematic expansion method indeed supports this assumption.

The heat locally generated by viscous heating effects is
redistributed by convection and thermal conduction, and partially compensated by
collisional dissipation.
A systematic expansion method to be presented in this paper works
when thermal conduction plays a dominant role in the reduction of 
temperature inhomogeneities through the redistribution and compensation
of heat.
The conditions necessary for this situation to be realized will 
be examined below.

We shall first discuss a typical size of temperature and density inhomogeneities when effects of convection and collisional dissipation are small compared
with thermal conduction effects, so that they may be neglected.
Denoting a typical size and length scale of the scaled flow field
are given by $\tilde{U}$ and $l_u$, respectively,
the viscous heating term in Eq.(\ref{stemperature}) is given, in order
of magnitude, by
\begin{eqnarray}
\frac{8}{d}D_{\perp}
\frac{\eta^{*}(\tilde{n})}{\tilde{n}}\sqrt{\tilde{T}}
\left(\tilde{\mbox{\sf D}}:\tilde{\mbox{\sf D}}\right)
&\simeq&
\frac{4}{d}D_{\perp}L^{-2} \tilde{U}^2\left(1+O(\Delta \tilde{T},\Delta \tilde{n})\right)
,
\label{viscousheating}
\end{eqnarray}
where $\Delta \tilde{T}$ and $\Delta \tilde{n}$ are a typical
size of $\tilde{T}-1$ and a typical size of $\tilde{n}-1$,
respectively, and assumed to be small. We shall show later that this
assumption is valid for $\tilde{U}^{2}\mayaprl 1$.
The thermal conduction term in Eq.(\ref{stemperature})
is similarly given by
\begin{eqnarray}
\frac{\bar{c}_p}{c_v}D_T\frac{1}{\tilde{n}}\nabla\cdot\left(\kappa^{*}(\tilde{n})\sqrt{\tilde{T}}\nabla \tilde{T}\right)
\simeq
\frac{\bar{c}_p}{c_v}D_T l_T^{-2}\Delta \tilde{T}
\left(1+O(\Delta\tilde{T},\Delta\tilde{n})\right)
,
\label{thermalconduction}
\end{eqnarray}
where $l_T$ is a typical length scale of temperature inhomogeneities.
Because the temperature inhomogeneities are induced by
viscous heating, which is proportional to the square of
$\nabla \tilde{{\bf u}}$,
$l_T$ can be approximately related to $l_u$ by
$l_T\sim l_u/2$.
The typical size of the inhomogeneous part of the scaled temperature
$\Delta \tilde{T}$ is then obtained from
the balance between these two terms (\ref{viscousheating}) and (\ref{thermalconduction}), i.e.,
\begin{eqnarray}
\Delta \tilde{T}
\simeq
\frac{c_v}{\bar{c}_p}\frac{P_r}{d}  \tilde{U}^2
\left(1+O(\Delta\tilde{T},\Delta\tilde{n})\right)
,
\label{tempscale}
\end{eqnarray}
where $P_r=D_{\perp}/D_{T}$ is the Prandtl number.
A typical size of the inhomogeneous part of the scaled density
can be estimated as follows:
In the asymptotic state we may assume that the pressure is homogeneous
as a consequence of the mechanical balance. Hence
the typical size of the inhomogeneous part of the scaled density
$\Delta \tilde{n}$ is related to $\Delta \tilde{T}$ by
\begin{eqnarray}
\Delta \tilde{n}
&=&
\bar{T}\bar{\beta}_{B}\Delta\tilde{T}
=\bar{\beta}_0\Delta\tilde{T}
,
\end{eqnarray}
where the constant $\bar{\beta}_B$ is the bulk expansion coefficient;
its definition is given in Appendix A. 
It is shown in Appendix A that for inelastic hard sphere fluids 
$\bar{\beta}_0=\bar{T}\bar{\beta}_B$ is a constant independent of $\bar{T}$,
and in two and three dimensions
$\bar{\beta}_0$ is less than $1$, i.e.,
$\Delta \tilde{n}<\Delta \tilde{T}$.

Values of $(c_v/\bar{c}_p)(P_r/d)$ for elastic hard spheres of
various densities are listed in Table~1.
As shown in Table~1, the factor $(c_v/\bar{c}_p)(P_r/d)$ is a small quantity
and hence $\Delta \tilde{T}\ll 1$ if $\tilde{U}^{2}\mayaprl 1$.
This result justifies the approximation used in Ref~\cite{santafe},
in which density and temperature inhomogeneities are neglected,
for a finite size of shear flow profile with $\tilde{U}^{2}\mayaprl 1$ in the asymptotic state.

\begin{table}
\begin{tabular}{|l|c|c|c|}
\hline
 & $P_r$ &  $\bar{c}_p/c_v$ & $(c_v/\bar{c}_p)(P_r/d)$\\
\hline
dilute limit (2D)& $0.5$ & $2$ & $0.12$\\
\hline
dilute limit (3D)& $0.67$ & $5/3$ & $0.13$\\
\hline
Enskog formula (2D, $\phi=0.1$) & $0.46$ & $2.0$ & $0.11$ \\
\hline
Enskog formula (3D, $\phi=0.1$) & $0.60$ & $1.7$ & $0.12$ \\
\hline
Enskog formula (2D, $\phi=0.4$) & $0.44$ & $2.2$ & $0.11$ \\
\hline
Enskog formula (3D, $\phi=0.4$) & $0.89$ & $2.4$ & $0.12$ \\
\hline
\end{tabular}
\caption{Values of $P_r$, $\bar{c}_p/c_v$ and $(c_v/\bar{c}_p)(P_r/d)$
given by the Boltzmann theory (dilute limit) and by the Enskog theory
(Enskog formula) for elastic hard spheres in two dimensions (2D) and in three dimensions (3D).
Two values of the packing fraction, $\phi=0.1$ and $0.4$,
are used for the Enskog formula.}
\end{table}

The fact that the inhomogeneous part of the scaled temperature and density
is small, and is proportional to the Prandtl number
motivates to expand the stationary solutions in the Prandtl
number, expecting that it will eventually lead an expansion in
the small parameter $(c_v/\bar{c}_p)(P_r/d)$,
which characterizes the small magnitude of density and temperature inhomogeneities
when $\tilde{U}^{2}\mayaprl 1$.
It will be shown that up to a part of the second approximation studied in this paper, this is indeed the case.

In Sec.4, we will discuss the range of validity of
this method and give a system size above which our
analysis breaks down.
It is noteworthy that temperature profile produced by the combination of
viscous hating effects and
thermal conduction effects appears also in flows of classical fluids such as the compressible
Couette flow for a fluid confined between isothermal or adiabatic walls~\cite{panton}.

Finally, let us consider the conditions under which
thermal conduction may be regarded as
dominant in comparison with convection and collisional dissipation.
The condition that the convection term in Eq. (\ref{stemperature}),
$l_0\tilde{\bf u}\cdot\nabla\tilde{T}\sim l_0 \tilde{U}l_T^{-1}\Delta\tilde{T}$, is negligible compared with the thermal conduction term yields,
\begin{eqnarray}
\frac{l_0 \tilde{U}l_T^{-1}\Delta\tilde{T}}
{\frac{\bar{c}_p}{c_v}D_T l_T^{-2}\Delta \tilde{T}}
=\left(\frac{c_v}{\bar{c}_p}\frac{P_r}{d}\right)
\frac{d}{2}R_e
\ll 1
.
\label{conv-thermcond}
\end{eqnarray}
Here $R_e$ is the Reynolds number defined by $R_e\equiv
U l_u/\nu$, where $U=\bar{v}_0\tilde{U}$ is a typical size
of the flow field (not scaled).
Smallness of the parameter $(c_v/\bar{c}_p)(P_r/d)$ 
suggests that this condition is fulfilled if
\begin{eqnarray}
\frac{d}{2}R_e \mayaprl 1
.
\label{conv-thermcond2}
\end{eqnarray}
The inhomogeneous part of the collisional
dissipation term, which contributes to reduction of inhomogeneities
in the temperature field, is in order of magnitude given by
$\gamma_0\Delta\tilde{T}$. The condition that this term is small
compared with the thermal conduction term yields,
\begin{eqnarray}
\frac{\gamma_0\Delta\tilde{T}}
{\frac{\bar{c}_p}{c_v}D_T l_T^{-2}\Delta \tilde{T}}
=\left(\frac{c_v}{\bar{c}_p}\frac{P_r}{d}\right)
\frac{d}{4}\frac{\gamma_0}{D_{\perp}l_u^{-2}}
\ll 1
.
\label{coldisp-thermcond}
\end{eqnarray}
Since $(c_v/\bar{c}_p)(P_r/d)\ll 1$, this condition is fulfilled
if
\begin{eqnarray}
\frac{d}{4}\frac{\gamma_0}{D_{\perp}l_u^{-2}}\mayaprl 1
.
\label{coldisp-thermcond2}
\end{eqnarray}

For the stationary solutions with the properties (i)-(v), 
which are to be studied in this paper, however, 
the property (iv) and the assumption of homogeneous pressure
suggest that temperature are inhomogeneous
only in directions perpendicular
to the direction of flow; hence the convection term
identically vanishes.
Therefore the condition (\ref{conv-thermcond2}) is not required
for this type of solutions.
Besides, it will be shown in Sec.3.4
that for solutions with the properties (i)-(v),
the condition (\ref{coldisp-thermcond2})
is fulfilled if $\tilde{U}^2\mayaprl 1$.

On the basis of the consideration denoted above, we expand
the stationary solution in the Prandtl number as follows:
\begin{eqnarray}
\tilde{n}&=& 1+\epsilon\tilde{n}^{(1)}+\epsilon ^2 \tilde{n}^{(2)}+\cdots
,
\nonumber\\
\tilde{{\bf u}}&=&
\tilde{{\bf u}}^{(0)}+\epsilon \tilde{{\bf u}}^{(1)}+\epsilon^2\tilde{{\bf u}}^{(2)}
+\cdots
,
\nonumber\\
\tilde{T}&=& 1 + \epsilon \tilde{T}^{(1)} + \epsilon^2 \tilde{T}^{(2)} + \cdots
.
\label{expansion}
\end{eqnarray}
Here we have introduced a parameter $\epsilon$ that is of $O(P_r)$.

We require that the conservation of the total number and
the total momentum, $N=\int d{\bf r} \,n$ and ${\bf 0}=\int d{\bf r} \,n{\bf u}$,
is satisfied in each order in $\epsilon$. This gives rise to the following conditions:
\begin{eqnarray}
0&=&\int d{\bf r}\,\tilde{n}^{(l)}\hspace{1cm}\mbox{for $l=1,2,\cdots$}
,
\label{c1}
\\
{\bf 0}&=&\int d{\bf r}\,\tilde{{\bf u}}^{(0)},\hspace{0.5cm}
{\bf 0}=\int d{\bf r}\left(\tilde{{\bf u}}^{(1)}+\tilde{n}^{(1)}\tilde{{\bf u}}^{(0)}\right),\hspace{0.5cm}
\cdots
.
\label{c2}
\end{eqnarray}
In addition, rewriting the definition of the global temperature as
$1=(1/V)\int d{\bf r} \,\tilde{n}\tilde{T}$ and requiring that this
relation is satisfied in each order in $\epsilon$, we get the following conditions
for the scaled temperature:
\begin{eqnarray}
0&=&\int d{\bf r}\,\tilde{T}^{(1)},\hspace{0.5cm}
0=\int d{\bf r}\left(\tilde{n}^{(1)}\tilde{T}^{(1)}+\tilde{T}^{(2)}\right),
\hspace{0.5cm}\cdots
.
\label{c3}
\end{eqnarray}

We substitute (\ref{expansion}) into the hydrodynamic equations
(\ref{sdensity})-(\ref{stemperature})
with the vanishing time derivatives
\begin{eqnarray}
\tilde{{\bf u}}\cdot \nabla \tilde{n}
&=&
-\tilde{n}\,\nabla\cdot
\tilde{{\bf u}}
,
\label{sdensitystat2}
\\
\tilde{n}\tilde{{\bf u}}\cdot l_0\nabla \tilde{{\bf u}}
&=&
-\frac{l_0}{2}\nabla \tilde{p}
+2D_{\perp}
\nabla\cdot\left(\eta^{*}(\tilde{n})
\sqrt{\tilde{T}} \,\tilde{\mbox{\sf D}}\right)
+\frac{\bar{\zeta}}{\bar{\rho}\bar{\omega}}
\nabla
\left(\zeta^{*}(\tilde{n})\sqrt{\tilde{T}}\nabla\cdot\tilde{{\bf u}}\right)
\nonumber\\
&-&
\frac{1}{2}\left(\partial_\tau \ln
\bar{T}\right)\tilde{n}\tilde{{\bf u}}
,
\label{svelocitystat2}
\\
\tilde{n}\tilde{{\bf u}}\cdot l_0\nabla \tilde{T}
&=&
-\frac{2}{d}\tilde{p}l_0\nabla\cdot \tilde{{\bf u}}
+\epsilon^{-1} \frac{\bar{c}_p}{c_v}D_T
\nabla\cdot\left(\kappa^{*}(\tilde{n})\sqrt{\tilde{T}}\nabla
\tilde{T}\right)
\nonumber\\
&+& \frac{8}{d}D_{\perp}
\eta^{*}(\tilde{n})\sqrt{\tilde{T}}
\left(\tilde{\mbox{\sf D}}:\tilde{\mbox{\sf D}}\right)
+\frac{4}{d}\frac{\bar{\zeta}}{\bar{\rho}\bar{\omega}}
\zeta^{*}(\tilde{n})\sqrt{\tilde{T}}
(\nabla\cdot\tilde{{\bf u}})^2
\nonumber\\
&-&
2\gamma_0\omega^{*}(\tilde{n})\tilde{n}\tilde{T}^{3/2}
-\left(\partial_{\tau} \ln \bar{T}\right)\tilde{n}\tilde{T}
.
\label{stemperaturestat2}
\end{eqnarray}
Here both sides of Eqs. (\ref{svelocity}) and (\ref{stemperature})
are multiplied by $\tilde{n}$ for later convenience.
We put $\epsilon^{-1}$ in front of the thermal conduction term,
because $P_r^{-1}$ appears there by rescaling collision time $\tau$
by $L^2/D_{\perp}$ and by rescaling space $r$ by $L$.
We omit this procedure since this rescaling
of variables by constants does not change the final result.

Collecting terms of the same order, we obtain a set of equations
that determine a stationary solution of the scaled hydrodynamic
variables in each order in $\epsilon$.
The smallness parameter $\epsilon$ is set equal to $1$ in the end
of calculation.

\subsection{Zeroth approximation}
The terms of $O(\epsilon^{-1})$ gives an identity $0=0$.
The zeroth approximation, $\tilde{{\bf u}}^{(0)}$, is determined
by the terms of $O(\epsilon^{0})$, i.e.
\begin{eqnarray}
0 &=& \nabla\cdot\tilde{{\bf u}}^{(0)}
,
\label{incomp}
\\
\tilde{{\bf u}}^{(0)}\cdot l_0\nabla \tilde{{\bf u}}^{(0)}
&=&
D_{\perp}
\nabla^2 \tilde{{\bf u}}^{(0)}
-
\frac{1}{2}\left(\partial_\tau \ln
\bar{T}\right)^{(0)}\tilde{{\bf u}}^{(0)}
,
\label{zerou}
\end{eqnarray}
where
\begin{eqnarray}
\left(\partial_\tau \ln \bar{T}\right)^{(0)}
&=&
\frac{1}{V}\int d{\bf r}\left[\,
\frac{4}{d}D_{\perp}\,
(\nabla\tilde{{\bf u}}^{(0)})^{\dagger}:
\nabla\tilde{{\bf u}}^{(0)}
-2\gamma_0
\,\right]
,
\label{zeroeqgtemp}
\end{eqnarray}
where $(\nabla \tilde{{\bf u}}^{(0)})^{\dagger}_{\alpha\beta}=\partial_{\beta}\tilde{u}_{\alpha}^{(0)}$.
We show in Appendix B that if $k_0>k^{*}_{\perp}$,
the nonlinear equation (\ref{zerou}) has only the trivial solution
$\tilde{{\bf u}}^{(0)}={\bf 0}$.
Here $k_0=2\pi/L$ is the smallest wave number of the system and
$k^{*}_{\perp}=\sqrt{\gamma_0/D_{\perp}}$ is the critical wave number
of the instability in shear modes, predicted in a linear stability
analysis~\cite{goldhirsch,goldhirsch2,mcnamara,deltour-barrat,brey1,vn-er-pre}.
In this case all higher order corrections vanish, suggesting that
the system is at large times in a homogeneous cooling state
that follows Haff's law~\cite{haff},
$
\left(\partial_\tau \ln \bar{T}\right)
=
\left(\partial_\tau \ln \bar{T}\right)^{(0)}
=
-2\gamma_0
$,
independently of the initial state of the system.

Hereafter, we shall consider only the non-trivial case $k_0<k^{*}_{\perp}$.
If $k_0<k^{*}_{\perp}$,
there is a large set of solutions~\cite{santafe} that satisfy Eqs. (\ref{incomp})-(\ref{zeroeqgtemp}).
In this paper we do not pursue to find all possible solutions
that satisfy Eqs. (\ref{incomp})-(\ref{zeroeqgtemp}). We show that there is indeed a solution
that corresponds to the flow profile observed in
simulations~\cite{goldhirsch,goldhirsch2,deltour-barrat,obne,soto,santafe,mcnamara-young}
with a period of the system size $L$.
An intuitive explanation of the flow profile with the period $L$ is
that this is the mode which shows the slowest hydrodynamic decay~\cite{goldhirsch,goldhirsch2}.
Assuming the flow profile on the basis of the observations,
we concentrate on a quantitative description of the amplitude of the flow profile,
which eventually leads us to a theoretical prediction of the amplitude of
inhomogeneous temperature and density profile.

One can show that the solutions of Eqs. (\ref{incomp})-(\ref{zeroeqgtemp}) with a period
of the system size $L$ have to
be a flow that is inhomogeneous in directions which are perpendicular to
the direction of the flow [see Appendix C],
i.e.
\begin{eqnarray}
\tilde{{\bf u}}^{(0)}({{\bf r}})
=
\sum_{\alpha=1}^{d'}\sum_{\beta=d'+1}^{d}
A_{\alpha \beta}\hat{{\bf e}}_{\beta}\cos(k_0 r_{\alpha}+\theta_{\alpha\beta})
,
\label{sold2}
\end{eqnarray}
where we have chosen $\alpha=1,2,\cdots,d'$ ($d'\ge 1$) 
as to be directions in which
$\tilde{{\bf u}}^{(0)}$ is inhomogeneous, and $\beta=d'+1,\cdots,d$ as
the remaining directions.
$\hat{{\bf e}}_{\beta}$ is a unit vector in the direction $\beta$,
and $\theta_{\alpha\beta}$ is an arbitrary phase factor.
The amplitudes $A_{\alpha\beta}$ are real numbers that satisfy the following conditions.
First, from Eq.~(\ref{incomp}), we get $A_{\alpha\alpha}=0$.
Second, if $k_0<k^{*}_{\perp}$, Eq. (\ref{zerou}) give rise to
a relation~\cite{santafe}:
\begin{eqnarray}
A_{0}^2
\equiv
\sum_{\alpha=1}^{d'}\sum_{\beta=d'+1}^{d}
\,A_{\alpha \beta}^2
=
\frac{d(\gamma_0-D_{\perp} k_0^2)}{D_{\perp} k_0^2}
.
\label{condit1}
\end{eqnarray}
The proof of the results (\ref{sold2}) and (\ref{condit1}) are given in Appendix C. From Eqs. (\ref{sold2}) and (\ref{condit1}) we obtain a typical size of
the scaled flow field
$\tilde{U}= [(1/V)\int d{\bf r}|\tilde{\bf u}^{(0)}({\bf r})|^{2}]^{1/2}
=A_0/\sqrt{2}$.

The achievement of one of the stationary solutions
can be viewed as a process of spontaneous symmetry breaking~\cite{santafe}.
Good quantitative agreement about the amplitude $A_0$ between the theoretical
prediction (\ref{condit1}) and the result of MD simulation has been shown in
Ref.~\cite{santafe}.

Substituting the solution (\ref{sold2}) into Eq. (\ref{zeroeqgtemp}), we get
\begin{eqnarray}
\left(\partial_\tau \ln \bar{T}\right)^{(0)}
&=&
-2D_{\perp}k_0^2
.
\label{0slope}
\end{eqnarray}

\subsection{First approximation}
First order correction $\tilde{T}^{(1)}$ can be obtained from the terms of
$O(\epsilon^{0})$,
\begin{eqnarray}
\nabla^2 \tilde{T}^{(1)}
&=&
-\frac{c_v}{\bar{c}_p}
\frac{4}{d}P_r\biggl\{
(\nabla\tilde{{\bf u}}^{(0)})^{\dagger}:
\nabla\tilde{{\bf u}}^{(0)}
-
\frac{1}{V}\int d{\bf r}\,
(\nabla\tilde{{\bf u}}^{(0)})^{\dagger}:
\nabla\tilde{{\bf u}}^{(0)}
\biggr\}
,
\label{tfirst}
\end{eqnarray}
where we have used a relation
$\nabla\tilde{{\bf u}}^{(0)}:\nabla\tilde{{\bf u}}^{(0)}=0$,
which is obtained from Eqs. (\ref{incomp}) and (\ref{zerou}).
The relation (\ref{tfirst}) manifests that the temperature inhomogeneities
are excited by the inhomogeneous part of the viscous heating term.
Substituting the expression (\ref{sold2}) and solving the Laplace equation
under the condition (\ref{c3}), we get
\begin{eqnarray}
\tilde{T}^{(1)}({\bf r})
&=&
-\frac{c_v}{\bar{c}_p}\frac{P_r}{2d}
\sum_{\alpha=1}^{d'}
A_{\alpha}^{'2} \cos\left[2\left(k_0 r_{\alpha}+\theta'_{\alpha}\right)\right]
,
\label{temp1}
\end{eqnarray}
where $A'_{\alpha}$ and $\theta'_{\alpha}$ are defined by
$A_{\alpha}^{'2} \cos\left[2\left(k_0 r_{\alpha}+\theta'_{\alpha}\right)\right]
=
\sum_{\beta=d'+1}^{d}
A_{\alpha\beta}^2 \cos\left[2\left(k_0 r_{\alpha}+\theta_{\alpha\beta}\right)\right]$.
It can be shown that if $d\ge 3$, there exist solutions with the constants
$A_{\alpha\beta}$ and $\theta_{\alpha\beta}$ 
with vanishing $A^{'2}_{\alpha}$.
However, in the following analysis we will disregard these solutions
because, to the best of our knowledge, the asymptotic states
corresponding to these solutions have not been reported.
Then, we will chose $\theta_{\alpha}'$ so that
$A_{\alpha}^{'2} = \sum_{\beta=d'+1}^{d}
A_{\alpha\beta}^2 \cos\left[2\left(\theta_{\alpha\beta}-\theta_{\alpha}'\right)\right]$
becomes positive.

The first approximation of the scaled density and flow field can be
obtained from the following relation given by the terms of $O(\epsilon^1)$:
\begin{eqnarray}
\tilde{{\bf u}}^{(0)}\cdot \nabla \tilde{n}^{(1)}
&=&
-\nabla\cdot
\tilde{{\bf u}}^{(1)}
,
\label{1sdensitystat}
\end{eqnarray}
\begin{eqnarray}
\lefteqn{
\tilde{{\bf u}}^{(0)}\cdot l_0\nabla \tilde{{\bf u}}^{(1)}
+
\tilde{{\bf u}}^{(1)}\cdot l_0\nabla \tilde{{\bf u}}^{(0)}
}\nonumber\\
&=&
-\frac{l_0}{2}\left[\left(\frac{\partial \tilde{p}}{\partial \tilde{n}}\right)_0
\nabla \tilde{n}^{(1)}
+
\left(\frac{\partial \tilde{p}}{\partial \tilde{T}}\right)_0
\nabla \tilde{T}^{(1)}
\right]
\nonumber\\
&+&
D_{\perp}\nabla\cdot
\left[\left(\left(\frac{\partial \eta^{*}}{\partial \tilde{n}}\right)_0
\tilde{n}^{(1)}+\frac{1}{2}\tilde{T}^{(1)}\right)
\left(\nabla\tilde{{\bf u}}^{(0)}+(\nabla\tilde{{\bf u}}^{(0)})^{\dagger}\right)
\right]
\nonumber\\
&+&
D_{\perp}\nabla\cdot
\left[
\nabla\tilde{{\bf u}}^{(1)}+(\nabla\tilde{{\bf u}}^{(1)})^{\dagger}
-\frac{2}{d}U(\nabla\cdot\tilde{{\bf u}}^{(1)})
\right]
+
\frac{\bar{\zeta}}{\bar{\rho}\bar{\omega}}
\nabla\nabla\cdot\tilde{{\bf u}}^{(1)}
\nonumber\\
&-&
\frac{1}{2}\left(\partial_\tau \ln \bar{T}\right)^{(0)}
\left(\tilde{n}^{(1)}\tilde{{\bf u}}^{(0)}+\tilde{{\bf u}}^{(1)}\right)
-
\frac{1}{2}\left(\partial_\tau \ln \bar{T}\right)^{(1)}\tilde{{\bf u}}^{(0)}
,
\label{1svelocitystat}
\end{eqnarray}
where $(\cdots)_0$ represents that $(\cdots)$ is evaluated
in the homogeneous state, i.e. $n=\bar{n}$ and $T=\bar{T}$;
$\left(\partial_\tau \ln \bar{T}\right)^{(0)}$ is given by Eq. (\ref{0slope}) and
\begin{eqnarray}
\left(\partial_\tau \ln \bar{T}\right)^{(1)}
&=&
\frac{1}{V}\int d{\bf r}\biggl[\,
\frac{4}{d}D_{\perp}
\left(\left(\frac{\partial \eta^{*}}{\partial \tilde{n}}\right)_0
\tilde{n}^{(1)}+\frac{1}{2}\tilde{T}^{(1)}\right)
(\nabla\tilde{{\bf u}}^{(0)})^{\dagger}:\nabla\tilde{{\bf u}}^{(0)}
\nonumber\\
&+&
\frac{8}{d}D_{\perp}
(\nabla\tilde{{\bf u}}^{(0)})^{\dagger}:\nabla\tilde{{\bf u}}^{(1)}
\biggr]
.
\label{1eqgtemp}
\end{eqnarray}
These equations can be solved analytically for
$\tilde{n}^{'(1)}$ and $\tilde{{\bf u}}^{'(1)}$
that are spatially averaged over the directions $\beta$ ($\beta=d'+1,\cdots,d$),
in which $\tilde{{\bf u}}^{(0)}$ is homogeneous:
\begin{eqnarray}
\tilde{n}^{'(1)}=
\frac{1}{L^{d-d'}}\int\prod_{\beta=d'+1}^{d} d r_{\beta}\,\, \tilde{n}^{(1)}
,\hspace{0.5cm}
\tilde{{\bf u}}^{'(1)}=
\frac{1}{L^{d-d'}}\int\prod_{\beta=d'+1}^{d} d r_{\beta}\,\, \tilde{{\bf u}}^{(1)}
.
\end{eqnarray}
By taking the spatial average over the directions $\beta$ ($\beta=d'+1,\cdots,d$),
of both sides of Eqs. (\ref{1sdensitystat}) and
(\ref{1svelocitystat}), we get a closed set of equations for
$\tilde{n}^{'(1)}$ and $\tilde{\bf u}^{'(1)}$:
\begin{eqnarray}
0
&=&
\sum_{\alpha=1}^{d'}\partial_{\alpha}\tilde{u}_{\alpha}^{'(1)}
\equiv
\nabla'\cdot \tilde{{\bf u}}^{'(1)}
,
\label{eq1}
\\
0
&=&
-\frac{l_0}{2}\left[\left(\frac{\partial \tilde{p}}{\partial \tilde{n}}\right)_0
\partial_{\alpha} \tilde{n}^{'(1)}
+
\left(\frac{\partial \tilde{p}}{\partial \tilde{T}}\right)_0
\partial_{\alpha} \tilde{T}^{(1)}
\right]
+D_{\perp}\nabla^{'2}\tilde{u}_{\alpha}^{'(1)}
+D_{\perp}k_0^2 \tilde{u}_{\alpha}^{'(1)}
,
\label{eq2}
\\
\lefteqn{
\tilde{{\bf u}}^{'(1)}\cdot l_0 \nabla'\tilde{u}^{(0)}_{\beta}
=
}
\nonumber\\
&+&
D_{\perp}\biggl[
\left(\left(\frac{\partial \eta^{*}}{\partial \tilde{n}}\right)_0
\nabla' \tilde{n}^{'(1)}+\frac{1}{2}\nabla' \tilde{T}^{(1)}\right)
\cdot\nabla' \tilde{u}_{\beta}^{(0)}
+
\left(\left(\frac{\partial \eta^{*}}{\partial \tilde{n}}\right)_0
\tilde{n}^{'(1)}+\frac{1}{2}\tilde{T}^{(1)}\right)
\nabla^{'2}\tilde{u}_{\beta}^{(0)}\biggr]
\nonumber\\
&+&
D_{\perp}\nabla^{'2} \tilde{u}_{\beta}^{'(1)}
+D_{\perp}k_0^2 \tilde{u}_{\beta}^{'(1)}
+D_{\perp}k_0^2\tilde{n}^{'(1)}\tilde{u}^{(0)}_{\beta}
-\frac{1}{2}\left(\partial_{\tau}\ln\bar{T}\right)^{(1)}
\tilde{u}_{\beta}^{(0)}
,
\label{eq3}
\end{eqnarray}
where $\alpha=1,\cdots,d'$ and $\beta=d'+1,\cdots,d$,
and $\nabla'$ has been defined by
$\nabla'=(\partial_1,\partial_2,\cdots,\partial_{d'})$.
Here Eq. (\ref{0slope}) has been used and
\begin{eqnarray}
\left(\partial_\tau \ln \bar{T}\right)^{(1)}
&=&
\frac{1}{L^{d'}}\int \prod_{\alpha=1}^{d'} d r_{\alpha}\,
\frac{4}{d}D_{\perp}\biggl[\,
\left(\left(\frac{\partial \eta^{*}}{\partial \tilde{n}}\right)_0
\tilde{n}^{'(1)}+\frac{1}{2}\tilde{T}^{(1)}\right)
(\nabla' \tilde{{\bf u}}^{(0)})^{\dagger}:\nabla' \tilde{{\bf u}}^{(0)}
\nonumber\\
&+&
2
(\nabla' \tilde{{\bf u}}^{(0)})^{\dagger}:\nabla' \tilde{{\bf u}}^{'(1)}
\biggr]
.
\nonumber\\
\label{1eqgtempspecial}
\end{eqnarray}
The equation (\ref{eq2}) can be solved without any difficulty.
Combination of Eqs. (\ref{eq1}) and (\ref{eq2}) yields
\begin{eqnarray}
0=
\left(\frac{\partial \tilde{p}}{\partial \tilde{n}}\right)_0
\nabla^{'2} \tilde{n}^{'(1)}
+
\left(\frac{\partial \tilde{p}}{\partial \tilde{T}}\right)_0
\nabla^{'2} \tilde{T}^{(1)}
.
\end{eqnarray}
Because of the periodic boundary conditions and the conditions (\ref{c1})
and (\ref{c3}), we obtain
\begin{eqnarray}
\tilde{n}^{'(1)}({\bf r})
&=&
-
\frac{\left(\frac{\partial \tilde{p}}{\partial \tilde{T}}\right)_0}
{\left(\frac{\partial \tilde{p}}{\partial \tilde{n}}\right)_0}
\tilde{T}^{(1)}({\bf r})
=
-
\frac{\frac{1}{\bar{n}}\left(\frac{\partial p}{\partial T}\right)_0}
{\frac{1}{\bar{T}}\left(\frac{\partial p}{\partial n}\right)_0}
\tilde{T}^{(1)}({\bf r})
=
-\bar{\beta}_0\tilde{T}^{(1)}({\bf r})
.
\label{dens1}
\end{eqnarray}
This relation is consistent with
the pressure balance achieved in the asymptotic state.
Substituting Eqs. (\ref{temp1}) and (\ref{dens1}) into Eq. (\ref{eq2}),
we get
\begin{eqnarray}
0
=
D_{\perp}\nabla^{'2}\tilde{u}_{\alpha}^{'(1)}
+
D_{\perp}k_0^2\tilde{u}_{\alpha}^{'(1)}
.
\label{ualpha}
\end{eqnarray}
This relation suggests in terms of the Fourier modes
$\tilde{u}^{'(1)}_{\alpha \,{\bf k}}=0$ if $|{\bf k}|\ne k_0$;
hence $\tilde{u}_{\alpha}^{'(1)}$ ($\alpha=1,\cdots,d'$) is given by
\begin{eqnarray}
\tilde{u}_{\alpha}^{'(1)}({\bf r})
=
\sum_{\alpha'=1}^{d'}C_{\alpha'\alpha}\cos(k_0 r_{\alpha'}+\zeta_{\alpha'\alpha})
,
\label{ualphasol}
\end{eqnarray}
where the amplitudes $C_{\alpha'\alpha}$ are real numbers
that satisfy $C_{\alpha\alpha}=0$
because of Eq. (\ref{eq1}), and the phase factors $\zeta_{\alpha'\alpha}$ are also real.

Since $\tilde{T}^{(1)}$, $\tilde{n}^{'(1)}$ and $\tilde{u}^{'(1)}_{\alpha}$
($\alpha=1,\cdots,d'$) are now known, Eq. (\ref{eq3}) is an inhomogeneous
linear equation for $\tilde{u}^{'(1)}_{\beta}$ ($\beta=d'+1,\cdots,d$).
We will show in Appendix D that
the solubility conditions of Eq. (\ref{eq3}) is satisfied if and only if
\begin{eqnarray}
\theta_{\alpha\beta}&=&\theta_{\alpha}'\,\,\,\mbox{or}\,\,\,\theta_{\alpha}'+\pi
,
\hspace{1cm}
\sum_{\beta=d'+1}^{d}A_{\alpha\beta}^2=A_{\alpha}^{'2}=\frac{1}{d'}A_0^2
,
\label{solubilitycondition}
\end{eqnarray}
for $\alpha=1,\cdots,d'$ and $\beta=d'+1,\cdots,d$.
The freedom of the choice $\theta_{\alpha\beta}=\theta_{\alpha}'\mbox{ or }\theta_{\alpha}'+\pi$
can be absorbed into the definition of $A_{\alpha\beta}$; hence Eqs. (\ref{sold2})
and (\ref{temp1}) can be rewritten as
\begin{eqnarray}
\tilde{{\bf u}}^{(0)}({\bf r})
&=&
\sum_{\alpha=1}^{d'}\sum_{\beta=d'+1}^{d}
A_{\alpha\beta}\hat{{\bf e}}_{\beta}\cos(k_0 r_{\alpha}+\theta'_{\alpha})
,
\label{sold3}
\\
\tilde{T}^{(1)}({\bf r})
&=&
\frac{\hat{T}}{d'}\sum_{\alpha=1}^{d'}
\cos\left[2\left(k_0 r_{\alpha}+\theta'_{\alpha}\right)\right]
,
\label{rtemp1}
\end{eqnarray}
where $\hat{T}\equiv -(c_v/\bar{c}_p)(P_r/2d)A_0^2$.
According to the relation (\ref{dens1}), $\tilde{n}^{'(1)}$ is given by
\begin{eqnarray}
\tilde{n}^{'(1)}({\bf r})
&=&
\frac{\hat{n}}{d'}
\sum_{\alpha=1}^{d'}
\cos\left[2\left(k_0 r_{\alpha}+\theta'_{\alpha}\right)\right]
,
\label{rdens1}
\end{eqnarray}
where $\hat{n}=-\bar{\beta}_0\hat{T}$.
Once the solubility conditions are satisfied,
calculation of the solution of Eq. (\ref{eq3})
which satisfies the condition (\ref{c2})
is straightforward.
The final result is presented in Appendix E.

Finally, substituting Eqs. (\ref{sold3}), (\ref{rtemp1}), (\ref{rdens1})
and (\ref{u1}) into (\ref{1eqgtempspecial}), we obtain
the first order correction to $(\partial_{\tau}\ln\bar{T})$,
\begin{eqnarray}
\left(\partial_{\tau}\ln\bar{T}\right)^{(1)}
&=&
D_{\perp}k_0^2\frac{1}{d'}
\left[\left(1+\left(\frac{\partial\eta^*}{\partial \tilde{n}}\right)_0\right)\hat{n}
+\frac{1}{2}\hat{T}\right]
.
\label{1slope}
\end{eqnarray}
Hence, from Eqs. (\ref{0slope}) and (\ref{1slope}),
the exponent $2\gamma_a$ in Eq. (\ref{temptau}) is given by
\begin{eqnarray}
2\gamma_a
&=&
-\left[
\left(\partial_{\tau}\ln\bar{T}\right)^{(0)}+\left(\partial_{\tau}\ln\bar{T}\right)^{(1)}
\right]
\nonumber\\
&=&
2D_{\perp}k_0^2
-
\frac{c_v}{\bar{c}_p}\frac{P_r}{2d'}\left(\gamma_0-D_{\perp}k_0^2\right)
\left[\left(1+\left(\frac{\partial\eta^*}{\partial \tilde{n}}\right)_0\right)\bar{\beta}_0
-\frac{1}{2}\right]
.
\label{sfirst}
\end{eqnarray}
Notice that the first order correction to $2\gamma_a$ depends explicitly on the coefficient of restitution
$\alpha$ through $\gamma_0$.
The ratio of the energy per particle to the global temperature is independent of time
and can be calculated from Eq. (\ref{prop}) to order $\epsilon$:
\begin{eqnarray}
\frac{E}{\frac{d}{2}\bar{T}}
&=&
\frac{2}{d}\frac{1}{V}\int d{\bf r}\,
\left(
\tilde{{\bf u}}^{(0)2}+
2\tilde{{\bf u}}^{(0)}\cdot\tilde{{\bf u}}^{(1)}
+
\tilde{n}^{(1)}\tilde{{\bf u}}^{(0)2}
\right)+1
\nonumber\\
&=&
\frac{\gamma_0}{D_{\perp}k_0^2}\left\{
1+\frac{c_v}{\bar{c}_p}\frac{P_r}{4d'}
\frac{\left(\gamma_0-D_{\perp}k_0^2\right)}{D_{\perp}k_0^2}
\left[\left(1+\left(\frac{\partial\eta^*}{\partial \tilde{n}}\right)_0\right)\bar{\beta}_0
-\frac{1}{2}\right]
\right\}
,
\label{propconst}
\end{eqnarray}
where Eqs. (\ref{sold3}), (\ref{rdens1}) and (\ref{u1}) have been used.

Calculation in the higher order approximation is rather involved.
In Appendix F we discuss the second order correction to the scaled
temperature,
$\tilde{T}^{(2)}$, in order to illustrate how
excitation of modes with larger wave numbers
can be described in the higher order approximations.

\subsection{The range of validity of the expansion method}
A restriction to the range of validity of the expansion method
is given by the condition that the amplitude of the temperature inhomogeneities
$|\hat{T}|=(c_v/\bar{c}_p)(P_r/2d)A_0^2=(c_v/\bar{c}_p)(P_r/d)\tilde{U}^2$,
which is greater than the amplitude
of the density inhomogeneities $|\hat{n}|$ in two and three dimensions,
has to be much smaller
than $1$.
In Sec.3.1 we have seen that $(c_v/\bar{c}_p)(P_r/d)$
is a small parameter; therefore the above condition is fulfilled
if
\begin{eqnarray}
\tilde{U}^2=\frac{A_0^2}{2}
=\frac{d}{2}\frac{\gamma_0-D_{\perp}k_0^2}{D_{\perp}k_0^2}
\mayaprl 1
.
\label{u2small}
\end{eqnarray}
This condition can be rewritten as
\begin{eqnarray}
\gamma_0\mayaprl \left(1+\frac{2}{d}\right)D_{\perp}k_0^2
,
\label{u2small2}
\end{eqnarray}
and hence tells how small the system has to be in order
that our systematic expansion method describes well
the asymptotic state.

We have also seen in Sec.3.1 that the inhomogeneous part of the collisional
dissipation term, which contributes to reduction of inhomogeneities
in the temperature field, is negligible if the condition (\ref{coldisp-thermcond2})
is fulfilled. From the condition (\ref{u2small2}), however, we see
that if $\tilde{U}^2\mayaprl 1$ then the condition (\ref{coldisp-thermcond2})
is fulfilled.
This result is consistent with the fact that the inhomogeneous
part of the collisional dissipation term appears in the second
order correction to the scaled temperature, as shown in Appendix F.
Therefore the range of validity of our systematic expansion
method applied to a solution with the properties (i)-(v) is given only by
the condition $\tilde{U}^2\mayaprl 1$.

\section{Conclusion}
In this paper, a systematic expansion method
in the Prandtl number has been developed in order to
describe the asymptotic state of freely cooling granular fluids
contained in sufficiently small systems.
Although the Prandtl number of IHS fluids itself is not a
very small parameter, it eventually leads an expansion
in the small parameter $(c_v/\bar{c}_p)(P_r/d)$, which characterizes
the size of temperature and density inhomogeneities if
a typical size of the scaled flow field
$\tilde{U}=A_0/\sqrt{2}$ satisfies $\tilde{U}^2 \mayaprl 1$.
The expansion in the Prandtl number yields a set of inhomogeneous linear equations.
The zeroth approximation is a homogeneous state
where only the scaled flow field is allowed to have
finite amplitude. The resulting equations in the zeroth
approximation are in agreement with those equations for the stationary state
which are given in Ref.~\cite{santafe} with an approximation in which
density and temperature inhomogeneities are neglected.
It has been shown that inhomogeneities in density and temperature
are systematically described in the higher order approximation;
corrections to macroscopic quantities such as the energy per
particle and the global temperature caused by these inhomogeneities
are estimated up to the first approximation.

Our theory describes sufficiently small systems where thermal conduction
plays a more important role than convection and collisional dissipation.
As a consequence, the temperature inhomogeneities is determined mainly by
viscous heating induced by the flow profile, and thermal conduction.
Density profile is then determined
in such a way that pressure averaged over the directions of
the flow in the zeroth approximation
remains homogeneous.
If the size of system is so large that
thermal conduction as a diffusion process
is less important than collisional dissipation
for large scale fluctuations of temperature,
a nonlinear theory by Goldhirsch et al.~\cite{goldhirsch,goldhirsch2}
is more appropriate to describe temperature profile determined mainly by
viscous heating and collisional dissipation;
their theory describes also pressure imbalance that eventually
leads a clustering process.
Apparently the clustering process can not be described by our
method because 
the condition of small inhomogeneities in density and temperature
is necessary for the systematic expansion method to work well.

The condition that the amplitude of
temperature inhomogeneities
$|\hat{T}|=(c_v/\bar{c}_p)(P_r/2d)A_0^2=(c_v/\bar{c}_p)(P_r/d)\tilde{U}^2$
has to be much smaller
than $1$
yields a restriction on the system size $L$ for a given
packing fraction and the coefficient of restitution $\alpha$ of IHS.
Because $(c_v/\bar{c}_p)(P_r/d)\ll 1$ in two and three dimensions [see Table 1],
our theory is expected to work well for $\tilde{U}^2\mayaprl 1$;
this is the case that can not be described by those amplitude equations in Ref.~\cite{soto}
which are derived assuming $\tilde{U}^2\ll 1$.
One can show that our results expanded in the small parameter in Ref.~\cite{soto}, $\tilde{U}^2=A_0^2/2$,
coincide to order $A_0^2 P_r^2$ with the results in Ref.~\cite{soto} expanded in the Prandtl number $P_r$,
except for the indefinite constants in our theory, which can not be
fixed in the first approximation.


In this paper, stability of the stationary solutions
has not been discussed.
We have assumed the flow profiles that are observed in computer simulations
(the property (i) in Sec.1)
and concentrated on the results that can be deduced from this assumption,
such as the amplitude of the scaled flow field, the amplitude of the density and temperature inhomogeneities, and the decay exponent of the global
temperature and the energy per particle.
The stability analysis may eliminate solutions that are not observed
in computer simulations and therefore not discussed in this paper.

In this paper the systematic expansion method has been applied to
freely evolving granular fluids contained in small systems with
periodic boundary conditions; these systems have been studied
in the earlier works by computer simulations and by theories.
The asymptotic state of these systems is unphysical
because it is dominated
by the finite size effect due to periodic boundary conditions.
Our method of systematic expansion, however, is expected
to be useful to study
other practical problems
which occur in finite size systems with physical boundary conditions,
such as
Couette flow and flow under vibration,
if the dimensions of the system are sufficiently small such that
viscous heating and thermal conduction play dominant roles
in formation of density and temperature inhomogeneities.

\section{Acknowledgements}
I thank M. H. Ernst, R. Brito, R. Soto and M. Mareschal for helpful discussions.
Thanks are due to M. H. Ernst for reading
the manuscript and making useful suggestions.
I acknowledges support of the foundation "Fundamenteel
Onderzoek der Materie (FOM)", which is financially supported by the Dutch
National Science Foundation (NWO).

\setcounter{equation}{0}
\renewcommand{\theequation}{A.\arabic{equation}}

\appendix
\section{Bulk expansion coefficient for inelastic hard spheres}
In this appendix, we present the expression for the bulk expansion coefficient
$\beta_B$ for $d$-dimensional IHS fluids and show that 
$\beta_0\equiv T\beta_B<1$
in two and three dimensions.
The bulk expansion coefficient $\beta_B$ is defined by
\begin{eqnarray}
\beta_B=-\frac{1}{n}\left.\frac{\partial n}{\partial T}\right|_p
=\frac{1}{n}\frac{\left.\frac{\partial p}{\partial T}\right|_n}
{\left.\frac{\partial p}{\partial n}\right|_T}
.
\label{bulkexp}
\end{eqnarray}
The expression for the local pressure in Enskog's kinetic theory
has been given in Refs.~\cite{jenkins-richman,jenkins-richman2} for three-dimensional IHS fluids.
A generalization to $d$ dimensions has been given in Ref.~\cite{pa-tr-vn-er}:
\begin{eqnarray}
p = n\,T
\left(1+\frac{1+\alpha}{2}\chi\, n \sigma^{d}\frac{\Omega_d}{2 d}\right)
,
\label{pressure}
\end{eqnarray}
where $\Omega_d=2\pi^{d/2}/\Gamma(d/2)$ is the surface area of a $d$-dimensional unit sphere,
$\sigma$ is the diameter of the hard sphere,
and $\chi$ is the pair correlation function for two particles in contact.
In the Enskog approximation $\chi$
has the same dependence on the local density at the point of contact
as the equilibrium pair correlation function at contact
has on the equilibrium density.
Substituting the expression (\ref{pressure}) into Eq. (\ref{bulkexp}),
we get
\begin{eqnarray}
\beta_B=\frac{1}{T}\frac{1+\frac{1+\alpha}{2}\chi\, n \sigma^{d}\frac{\Omega_d}{2 d}}
{1+\frac{1+\alpha}{2}\chi \,n
\left(1+\frac{1}{\chi}\frac{\partial}{\partial n}\left(\chi\, n\right)\right)
\sigma^{d}\frac{\Omega_d}{2 d}}
.
\label{bulkexp2}
\end{eqnarray}
From this expression, we see that 
$\beta_0=T\beta_B$ is independent of T, and if
\begin{eqnarray}
\frac{\partial}{\partial n}\left(\chi\, n\right)>0
,
\label{positive}
\end{eqnarray}
then $\beta_0<1$.
The condition (\ref{positive}) is satisfied in three dimensions
for $\chi$ given by the Carnahan-Starling approximation~\cite{car-sta},
$\chi=(2-\phi)/2(1-\phi)^3$,
and in two dimensions for $\chi$ given by the Verlet-Levesque
approximation~\cite{ver-lev}, $\chi=(1-7\phi/16)/(1-\phi)^2$, where
$\phi=n \Omega_d (\sigma/2)^d/d$ is the packing fraction in
$d$ dimensions.

\section{Homogeneous cooling state}
In this appendix, we show that if $k>k^{*}_{\perp}$,
Eq. (\ref{zerou}) has only the trivial solution $\tilde{{\bf u}}^{(0)}={\bf 0}$.
We first take the inner product of $\tilde{{\bf u}}^{(0)}$ with Eq. (\ref{zerou})
and integrate both sides over the whole space:
\begin{eqnarray}
\int d{\bf r}\tilde{{\bf u}}^{(0)}\cdot\left(
\tilde{{\bf u}}^{(0)}\cdot l_0\nabla \tilde{{\bf u}}^{(0)}
\right)
&=&
D_{\perp}
\int d{\bf r}\tilde{{\bf u}}^{(0)}\cdot
\nabla^2 \tilde{{\bf u}}^{(0)}
-
\frac{1}{2}\left(\partial_\tau \ln
\bar{T}\right)^{(0)}\int d{\bf r}(\tilde{{\bf u}}^{(0)})^2
.
\label{integrate}
\end{eqnarray}
Using the relation
$\tilde{{\bf u}}^{(0)}\cdot(\tilde{{\bf u}}^{(0)}\cdot \nabla \tilde{{\bf u}}^{(0)})
=\tilde{{\bf u}}^{(0)}\cdot \nabla (\tilde{{\bf u}}^{(0)})^2/2$
and performing an integration by parts,
it can be shown that the left hand side of Eq. (\ref{integrate})
vanishes as a consequence of Eq. (\ref{incomp}) and the periodic boundary conditions.
Rewriting the right hand side in terms of Fourier modes
defined as $a_{{\bf k}}=\int d{\bf r}e^{-i{\bf k}\cdot{\bf r}}a({\bf r})$, we get
\begin{eqnarray}
0
=
\sum_{{\bf k}}
\left(\gamma_0-D_{\perp}k^2
-\frac{2}{d}D_{\perp}\,
\frac{1}{V^2}\sum_{{\bf k}_1}k_1^2
|\tilde{{\bf u}}^{(0)}_{{\bf k}_1}|^2
\right)
|\tilde{{\bf u}}^{(0)}_{{\bf k}}|^2
.
\label{fourierzerou}
\end{eqnarray}
We notice that if $k_0>k^{*}_{\perp}\equiv\sqrt{\gamma_0/D_{\perp}}$,
the inside of the parenthesis on the right hand side of Eq. (\ref{fourierzerou})
is negative for any ${\bf k}$.
Therefore, the equality is satisfied if and only if $\tilde{{\bf u}}^{(0)}={\bf 0}$.

\section{Flow profile given in \protect{(\ref{sold2})} }
The flow field $\tilde{{\bf u}}^{(0)}$ that consists only of
the Fourier modes with the smallest wave number $k_0$
is written in general as~\cite{santafe}
\begin{eqnarray}
\tilde{{\bf u}}^{(0)}({\bf r})
=\sum_{\alpha=1}^{d}\sum_{\beta=1}^{d}
A_{\alpha\beta}\hat{{\bf e}}_{\beta} \cos (k_0 r_{\alpha}+\theta_{\alpha\beta})
,
\label{soldA}
\end{eqnarray}
where the coefficients $A_{\alpha\beta}$ and $\theta_{\alpha\beta}$
are arbitrary real constants.
From Eq. (\ref{incomp}), we get $A_{\alpha\alpha}=0$.

For the flow field given by Eq. (\ref{soldA}), both sides
of Eq. (\ref{zerou}) have to vanish independently of each other, because
the right hand side of Eq. (\ref{zerou}) consists only of the Fourier modes with
the smallest wave number $k_0$, while
the left hand side of Eq. (\ref{zerou}) does not contain the smallest wave
number modes.
We first show that the condition
of the vanishing left hand side of Eq. (\ref{zerou}) leads to
the flow field (\ref{sold2}).
The condition of the vanishing left hand side can be written in terms of the Fourier
modes,
\begin{eqnarray}
{\bf 0}
&=&
\sum_{{\bf k}_1}\tilde{{\bf u}}^{(0)}_{{\bf k}-{\bf k}_1}\cdot
{\bf k}_1\,\tilde{{\bf u}}^{(0)}_{{\bf k}_1}
,
\label{conv0}
\end{eqnarray}
for any ${\bf k}$.
Because $\tilde{{\bf u}}^{(0)}_{{\bf k}}={\bf 0}$ if $|{\bf k}|\ne k_0$,
the non-trivial conditions are given only for the wave numbers ${\bf k}$
that satisfy
$|{\bf k}\mp {\bf k}_{0\alpha}|=k_0$ $(\alpha=1,2,\cdots,d)$,
or, equivalently, only for ${\bf k}={\bf k}_{0\alpha}\pm {\bf k}_{0\beta}$
$(\alpha=1,2,\cdots,d,\,\beta=1,2,\cdots,d)$.
Here ${\bf k}_{0\alpha}=k_0\hat{{\bf e}}_{\alpha}$ and
$\hat{{\bf e}}_{\alpha}$ is a unit vector in the direction $\alpha$.
For a given ${\bf k}={\bf k}_{0\alpha}+{\bf k}_{0\beta}$,
the condition (\ref{conv0}) is written as
\begin{eqnarray}
{\bf 0}
&=&
\tilde{{\bf u}}^{(0)}_{{\bf k}_{0\alpha}}\cdot
{\bf k}_{0\beta}\,\tilde{{\bf u}}^{(0)}_{{\bf k}_{0\beta}}
+
\tilde{{\bf u}}^{(0)}_{{\bf k}_{0\beta}}\cdot
{\bf k}_{0\alpha}\,\tilde{{\bf u}}^{(0)}_{{\bf k}_{0\alpha}}
,
\\
{\bf 0}
&=&
-\tilde{{\bf u}}^{(0)}_{{\bf k}_{0\alpha}}\cdot
{\bf k}_{0\beta}\,\tilde{{\bf u}}^{(0)*}_{{\bf k}_{0\beta}}
+
\tilde{{\bf u}}^{(0)*}_{{\bf k}_{0\beta}}\cdot
{\bf k}_{0\alpha}\,\tilde{{\bf u}}^{(0)}_{{\bf k}_{0\alpha}}
.
\label{ccc}
\end{eqnarray}
Substituting the Fourier transform of Eq. (\ref{soldA}),
$\tilde{{\bf u}}_{{\bf k}_{0\alpha}}=(V/2)\sum_{\beta=1}^{d}\hat{{\bf e}}_{\beta}B_{\alpha\beta}$,
where $B_{\alpha\beta}=A_{\alpha\beta}\exp(i\theta_{\alpha\beta})$,
into the above condition (\ref{ccc}), we obtain
\begin{eqnarray}
0=
B_{\alpha\beta}B_{\beta\gamma}
+
B_{\beta\alpha}B_{\alpha\gamma},
\label{bcond1}
\\
0=
B_{\alpha\beta}B_{\beta\gamma}^{*}
-
B_{\beta\alpha}^{*}B_{\alpha\gamma},
\label{condit2prime}
\end{eqnarray}
for any set of $(\alpha,\beta,\gamma)$.

This conditions imply the following properties of the complex matrix $B_{\alpha\beta}$.
First, considering the case when $\gamma=\alpha$ in Eq. (\ref{bcond1}), we get
\begin{eqnarray}
0=B_{\alpha\beta}B_{\beta\alpha}
,
\label{alphabeta}
\end{eqnarray}
for any $\alpha$ and $\beta$. Here we have used property that
$B_{\alpha\alpha}=0$.
Second, suppose $B_{\alpha\beta}\ne 0$. Then, $B_{\beta\alpha}=0$ from Eq. (\ref{alphabeta}).
Then, the condition (\ref{bcond1}) is reduced to $0=B_{\alpha\beta}B_{\beta\gamma}$.
Because $B_{\alpha\beta}\ne 0$, we get $B_{\beta\gamma}=0$ for any $\gamma$.
Then suppose $A_{\alpha\beta}\ne 0$ for a set of
$(\alpha,\beta)=(\mu_1,\nu_1),(\mu_2,\nu_2),\cdots,(\mu_m,\nu_m)$
and $A_{\alpha\beta}=0$ otherwise.
Let us label the directions in the subspace spanned by $(\mu_1,\mu_2,\cdots,\mu_m)$ as
$(1,2,\cdots,d')$, where $d'$ is the number of dimensions of the subspace.
Note that $d'\le m$ because some of $\mu_i$ may be the same direction.
Then one finds that any $\nu_i$ can be equal to none of $(1,2,\cdots,d')$,
because $B_{\nu_i\gamma}=A_{\nu_i\gamma}=0$ for any $\gamma$.
The condition (\ref{condit2prime}) gives no additional conditions.
Hence, we may conclude that
the direction of the flow, which is in the subspace spanned by $(\nu_1,\nu_2,\cdots,\nu_m)$
has to be perpendicular to the directions $(1,2,\cdots,d')$ in which the flow shows inhomogeneities.
Therefore the flow can be in general written as Eq. (\ref{sold2}).

Then it is straightforward to get the condition (\ref{condit1})
by substituting the expression (\ref{sold2})
into the condition of the vanishing right hand side of Eq. (\ref{zerou}).

\section{Solubility conditions of Eq.~\protect{(\ref{eq3})}}
Equation (\ref{eq3}) is the inhomogeneous linear equation
with respect to $\tilde{u}^{'(1)}_{\beta}$ and rewritten as
\begin{eqnarray}
{\cal L}\,\tilde{u}^{'(1)}_{\beta}=f_{\beta}
,
\label{lininhomog}
\end{eqnarray}
where
\begin{eqnarray}
{\cal L}\,\tilde{u}^{'(1)}_{\beta}
&=&
D_{\perp}\nabla^{'2}\tilde{u}^{'(1)}_{\beta}
+D_{\perp}k_0^2\tilde{u}^{'(1)}_{\beta}
-k_0^2\left[
\frac{1}{L^{d'}}\int\prod_{\alpha=1}^{d'} d r_{\alpha}\,
\tilde{{\bf u}}^{(0)}\cdot \tilde{{\bf u}}^{'(1)}
\right]
\tilde{u}^{(0)}_{\beta}
,
\end{eqnarray}
and
\begin{eqnarray}
f_{\beta}
&=&
\tilde{{\bf u}}^{'(1)}\cdot l_0 \nabla'\tilde{u}^{(0)}_{\beta}
-D_{\perp}k_0^2\tilde{n}^{'(1)}\tilde{u}_{\beta}^{(0)}
\nonumber\\
&-&
D_{\perp}\biggl[
\left(\left(\frac{\partial \eta^{*}}{\partial \tilde{n}}\right)_0
\nabla' \tilde{n}^{'(1)}+\frac{1}{2}\nabla' \tilde{T}^{(1)}\right)
\cdot\nabla' \tilde{u}_{\beta}^{(0)}
+
\left(\left(\frac{\partial \eta^{*}}{\partial \tilde{n}}\right)_0
\tilde{n}^{'(1)}+\frac{1}{2}\tilde{T}^{(1)}\right)
\nabla^{'2}\tilde{u}_{\beta}^{(0)}\biggr]
\nonumber\\
&+&
\frac{2}{d}D_{\perp}\left[
\frac{1}{L^{d'}}
\int\prod_{\alpha=1}^{d'} d r_{\alpha}\,
\left(\left(\frac{\partial \eta^{*}}{\partial \tilde{n}}\right)_0
\tilde{n}^{'(1)}+\frac{1}{2}\tilde{T}^{(1)}\right)
(\nabla' \tilde{{\bf u}}^{(0)})^{\dagger}:\nabla' \tilde{{\bf u}}^{(0)}
\right]
\tilde{u}_{\beta}^{(0)}
,
\label{ef}
\end{eqnarray}
where $\tilde{{\bf u}}^{(0)}$, $\tilde{T}^{(1)}$, $\tilde{n}^{(1)}$, and $\tilde{{\bf u}}^{(1)}$
are given by Eqs. (\ref{sold2}),(\ref{temp1}),(\ref{dens1}),(\ref{u1}) and (\ref{ualphasol}).
Let us define a Hilbert space $\left|a\right>$ to represent any function
$a_\beta({\bf r})$ ($\beta=d'+1,\cdots,d,\,{\bf r}=(r_1,r_2,\cdots,r_{d'}))$
by
\begin{eqnarray}
a_\beta({\bf r})=\left<{\bf r},\beta|a\right>
.
\end{eqnarray}
We define the scalar product of two vectors $\left|a\right>$ and $\left|b\right>$ by
\begin{eqnarray}
\left<a|b\right>=
\frac{1}{L^{d'}}\int\prod_{\alpha=1}^{d'} d r_{\alpha}\,
\sum_{\beta=d'+1}^{d}
a^*_{\beta}({\bf r})\,b_{\beta}({\bf r})
.
\end{eqnarray}
With this definition, the inhomogeneous linear equation (\ref{lininhomog}) can be written as
\begin{eqnarray}
{\cal L} |\tilde{u}^{'(1)}\bigr>=\left|f\right>
,
\label{lininhomog2}
\end{eqnarray}
and the linear operator ${\cal L}$ is Hermitian:
\begin{eqnarray}
\left<a|{\cal L}b\right>=\left<b|{\cal L}a\right>
.
\end{eqnarray}
The inhomogeneous linear equation (\ref{lininhomog2}) has a nontrivial solution
if and only if $\left|f\right>$ is orthogonal to the solutions $\bigl|\phi^l\bigr>$ of the homogeneous
problem ${\cal L}\,\bigl|\phi^{l}\bigr>=0$~\cite{den-krz}, i.e.
\begin{eqnarray}
\bigl<\phi^{l}|f\bigr>=0
.
\label{solubilityconditions}
\end{eqnarray}

One can show that the linearly independent solutions $|\phi^l\bigr>$ of the homogeneous problem are given by
\begin{eqnarray}
\phi_{c\beta}^l({\bf r})
&=&
\bigl<{\bf r},\beta|\phi_c^l\bigr>
=
\sum_{\alpha=1}^{d'}c_{\alpha\beta}^l\cos(k_0 r_{\alpha}+\theta_{\alpha\beta})
,
\label{solcos}
\\
\phi_{s\beta}^m({\bf r})
&=&
\bigl<{\bf r},\beta|\phi_s^m\bigr>
=
\sum_{\alpha=1}^{d'}s_{\alpha\beta}^m\sin(k_0 r_{\alpha}+\theta_{\alpha\beta})
,
\label{solsin}
\end{eqnarray}
where the coefficients $c_{\alpha\beta}^l$ $(l=1,2,\cdots,d'(d-d')-1)$
and $s_{\alpha\beta}^m$ $(m=1,2,\cdots,d'(d-d'))$ are real numbers.
The coefficients $c_{\alpha\beta}^l$ satisfy the condition
\begin{eqnarray}
\sum_{\alpha=1}^{d'}\sum_{\beta=d'+1}^{d}A_{\alpha\beta}c_{\alpha\beta}^l=0
.
\label{condforc}
\end{eqnarray}
We also require the orthogonality property
\begin{eqnarray}
\bigl<\phi_c^l|\phi_c^{l'}\bigr>
&=&
\frac{1}{2}
\sum_{\alpha=1}^{d'}\sum_{\beta=d'+1}^{d}
c_{\alpha\beta}^l c_{\alpha\beta}^{l'}=\delta_{l l'}
,
\label{cortho}
\\
\bigl<\phi_s^m|\phi_s^{m'}\bigr>
&=&
\frac{1}{2}
\sum_{\alpha=1}^{d'}\sum_{\beta=d'+1}^{d}
s_{\alpha\beta}^m s_{\alpha\beta}^{m'}=\delta_{m m'}
,
\label{orthocond}
\\
\bigl<\phi_c^l|\phi_s^{m}\bigr>
&=&
\bigl<\phi_s^m|\phi_c^{l}\bigr>
=0
.
\end{eqnarray}

The solubility conditions (\ref{solubilityconditions}),
$\bigl<\phi_c^{l}|f\bigr>=0$ and $\bigl<\phi_s^{m}|f\bigr>=0$,
yield
\begin{eqnarray}
\sum_{\alpha=1}^{d'}\sum_{\beta=d'+1}^{d}
A^{'2}_{\alpha}A_{\alpha\beta}c_{\alpha\beta}^l\cos[2(\theta'_\alpha-\theta_{\alpha\beta})]
&=&
0
,
\label{condc}
\end{eqnarray}
and
\begin{eqnarray}
\sum_{\alpha=1}^{d'}\sum_{\beta=d'+1}^{d}
A^{'2}_{\alpha}A_{\alpha\beta}s_{\alpha\beta}^m\sin[2(\theta'_\alpha-\theta_{\alpha\beta})]
&=&
0
,
\label{conds}
\end{eqnarray}
respectively. Here we have assumed $[(1+(\partial\eta^*/\partial\tilde{n})_0)\bar{\beta}_0-1/2]\ne 0$,
which is satisfied in general. In deriving Eq. (\ref{conds})
we have used the fact that $\tilde{u}_{\alpha{\bf k}}^{(1)}=0$ $(\alpha=1,\cdots d')$ if $|{\bf k}|\ne k_0$
as shown in Eq. (\ref{ualphasol}).

Because the coefficient $s_{\alpha\beta}^1$ is arbitrary, let us choose it as
\begin{eqnarray}
s_{\alpha\beta}^1
\propto
A_{\alpha\beta}\sin[2(\theta'-\theta_{\alpha\beta})]
.
\label{s1}
\end{eqnarray}
Substituting this expression into Eq. (\ref{conds}), we get
\begin{eqnarray}
\sum_{\alpha=1}^{d'}\sum_{\beta=d'+1}^{d}
A_{\alpha}^{'2}A_{\alpha\beta}^2\sin^2[2(\theta'_\alpha-\theta_{\alpha\beta})]
&=&
0
.
\label{conds2}
\end{eqnarray}
Because $A^{'2}_{\alpha} > 0$ by definition [see note below
Eq. (\ref{temp1})], the condition (\ref{conds2}) suggests that if $A_{\alpha\beta}\ne 0$, then
$(\theta_{\alpha\beta}-\theta'_{\alpha})$ is either $0,\pi/2,\pi$, or $3\pi/2$.
We note that if $A_{\alpha\beta}=0$ then the phase factor $\theta_{\alpha\beta}$ does not appear
in our theory [see, Eq. (\ref{sold2})].
If the condition (\ref{conds2}) is satisfied,
the conditions (\ref{conds}) for $s_{\alpha\beta}^m$ ($m=2,3,\cdots,d'(d-d')$)
are automatically satisfied.

The condition (\ref{condc}) has to be satisfied for a set of constants
$c^{l}_{\alpha\beta}$, which
satisfies Eqs. (\ref{condforc}) and (\ref{cortho}). This implies that
$A_{\alpha}^{'2}A_{\alpha\beta}\cos[2(\theta_{\alpha}'-\theta_{\alpha\beta})]
= \mbox{const.}\times A_{\alpha\beta}$, i.e.
\begin{eqnarray}
A_{\alpha}^{'2}\cos[2(\theta_{\alpha}'-\theta_{\alpha\beta})]
=
\mbox{const.}={\cal Z}
,
\label{const}
\end{eqnarray}
where the constant ${\cal Z}$ is independent of $\alpha$ and $\beta$.
The condition that the left
hand side of Eq. (\ref{const}) is independent of $\beta$ suggests that
$(\theta_{\alpha\beta}-\theta'_{\alpha})$ is either $0$ or $\pi$, or
$(\theta_{\alpha\beta}-\theta'_{\alpha})$ is either $\pi/2$ or $3\pi/2$.
Because of the definition of $\theta_{\alpha}'$ [see note below Eq. (\ref{temp1})], however,
only the former choice
\begin{eqnarray}
\theta_{\alpha\beta}-\theta'_{\alpha}=0 \mbox{ or }\pi
\label{phasetheta}
\end{eqnarray}
is allowed.
Then, from Eqs. (\ref{const}) and (\ref{phasetheta})
we get $A_{\alpha}^{'2}=\sum_{\beta=d'+1}^{d}A_{\alpha\beta}^2= {\cal Z}$.
Finally, the constant ${\cal Z}$ is determined by the condition (\ref{condit1}),
\begin{eqnarray}
A_0^2
=
\sum_{\alpha=1}^{d'}
A^{'2}_{\alpha}
={\cal Z}\,d'
.
\end{eqnarray}

\section{First order correction \protect{$\tilde{u}_{\beta}^{'(1)}$}}

First order correction $\tilde{u}_{\beta}^{'(1)}$ can be obtained
by solving Eq. (\ref{eq3}) under the solubility conditions (\ref{solubilitycondition}) and
the condition (\ref{c2}):
\begin{eqnarray}
\lefteqn{
\tilde{u}^{'(1)}_{\beta}({\bf r})
=
{\cal U}_1
\sum_{\alpha=1}^{d'}A_{\alpha\beta}\cos[3(k_0 r_{\alpha}+\theta'_{\alpha})]
}
\nonumber\\
&+&
{\cal U}_2
\sum_{\alpha=1}^{d'}\sum_{\alpha'=1 \atop \alpha'\ne\alpha}^{d'}A_{\alpha'\beta}
\left\{
\cos\left[k_0(2r_{\alpha}-r_{\alpha'})+2\theta'_{\alpha}-\theta'_{\alpha'}\right]
+
\cos\left[k_0(2r_{\alpha}+r_{\alpha'})+2\theta'_{\alpha}+\theta'_{\alpha'}\right]
\right\}
\nonumber\\
&+&
\frac{l_0}{D_{\perp}k_0}\sum_{\alpha=1}^{d'}A_{\alpha\beta}\tilde{u}_{\alpha}^{'(1)}({\bf r})
\sin(k_0 r_{\alpha}+\theta'_{\alpha})
\nonumber\\
&+&\sum_{\alpha=1}^{d'}E_{\alpha\beta} \cos(k_0 r_{\alpha}+\theta'_{\alpha})
+\sum_{\alpha=1}^{d'}F_{\alpha\beta} \sin(k_0 r_{\alpha}+\theta'_{\alpha})
,
\label{u1}
\end{eqnarray}
where the coefficients $E_{\alpha\beta}$ are real numbers that satisfy the condition
\begin{eqnarray}
\sum_{\alpha=1}^{d'}\sum_{\beta=d'+1}^{d}A_{\alpha\beta}E_{\alpha\beta}
=
{\cal U}_3
,
\end{eqnarray}
and the coefficients $F_{\alpha\beta}$ are arbitrary real numbers.
Here the constants ${\cal U}$'s are defined by
\begin{eqnarray}
{\cal U}_1
&=&
\frac{1}{16 d'}\left[\left(1-3\left(\frac{\partial\eta^*}{\partial \tilde{n}}\right)_0\right)\hat{n}
-\frac{3}{2}\hat{T}\right]
,
\\
{\cal U}_2
&=&
\frac{1}{8 d'}\left[\left(1-\left(\frac{\partial\eta^*}{\partial \tilde{n}}\right)_0\right)\hat{n}
-\frac{1}{2}\hat{T}\right]
,
\\
{\cal U}_3
&=&
\frac{d}{4 d'}
\left[\left(1+\left(\frac{\partial\eta^*}{\partial \tilde{n}}\right)_0\right)\hat{n}
+\frac{1}{2}\hat{T}\right]
+\frac{1}{4 d'}
\left(\left(\frac{\partial\eta^*}{\partial \tilde{n}}\right)_0\hat{n}
+\frac{1}{2}\hat{T}\right)
A_0^2
.
\end{eqnarray}
It should be noted that all of the ${\cal U}$'s are proportional to
the small parameter $(c_v/\bar{c}_p)(P_r/d)$.
The results (\ref{ualphasol}) and (\ref{u1}) show that $\tilde{{\bf u}}^{'(1)}$
has indefinite constants that can not be fixed in the first 
approximation.
We expect that they are fixed by the solubility conditions of the inhomogeneous
linear equations in the higher order approximations.
As shown in Eqs. (\ref{sfirst}) and (\ref{propconst}), however,
the expressions for macroscopic quantities such as the global temperature and
the energy per particle up to the first approximation
do not contain these indefinite constants.

\section{Second order correction $\tilde{T}^{(2)}$}
The purpose of this appendix is to illustrate how
excitation of modes with larger wave numbers
can be described in the higher order approximations.

The spatial average over the directions $\beta=d'+1,\cdots,d$
of both sides of the $O(\epsilon^1)$ equation
obtained from Eq. (\ref{stemperaturestat2}) yields
\begin{eqnarray}
\lefteqn{
\frac{\bar{c}_p}{c_v}D_{T}
\nabla^{'2}\tilde{T}^{'(2)}
=
-\frac{\bar{c}_p}{c_v}D_{T}
\nabla'\cdot\left[\left(\left(\frac{\partial \kappa^*}{\partial \tilde{n}}\right)_0
\tilde{n}^{'(1)}+\frac{1}{2}\tilde{T}^{(1)}\right)\nabla' \tilde{T}^{(1)} \right]
}
\nonumber\\
&-&
\frac{4}{d}D_{\perp}\left[\left(\frac{\partial \eta^*}{\partial \tilde{n}}\right)_0
\tilde{n}^{'(1)}+\frac{1}{2}\tilde{T}^{(1)}\right]
(\nabla'\tilde{{\bf u}}^{(0)})^{\dagger}:\nabla'\tilde{{\bf u}}^{(0)}
-
\frac{8}{d}D_{\perp}
(\nabla'\tilde{{\bf u}}^{(0)})^{\dagger}:\nabla'\tilde{{\bf u}}^{'(1)}
\nonumber\\
&+&
2\gamma_0\left[\left(\frac{\partial \omega^*\tilde{n}}{\partial\tilde{n}}\right)_0\tilde{n}^{'(1)}
+\frac{3}{2}\tilde{T}^{(1)}\right]
+
\left(\partial_{\tau}\ln\bar{T}\right)^{(0)}\left(\tilde{n}^{'(1)}+\tilde{T}^{(1)}\right)
+\left(\partial_{\tau}\ln\bar{T}\right)^{(1)}
,
\label{secondtemp}
\end{eqnarray}
where $\tilde{T}^{'(2)}$ is that second order correction
to the scaled temperature
which is spatially averaged over the directions $\beta=d'+1,\cdots,d$.
This is again the inhomogeneous linear equation (the Laplace equation)
for $\tilde{T}^{'(2)}$ and it can be shown
by substituting the result of the first approximation
that the solubility conditions of this equation are satisfied.
Because the general expression
for $\tilde{T}^{'(2)}$ is rather lengthy and complicated, we concentrate here only
on the case that the zeroth approximation $\tilde{{\bf u}}^{(0)}$ is
inhomogeneous only in one direction, say $\alpha=1$, i.e.
$d'=1$;
this is always true in two dimensions if $k_0<k_{\perp}^{*}$.
Then $\tilde{u}_1^{'(1)}=0$ and for $\beta=2,3,\cdots,d$,
\begin{eqnarray}
\tilde{u}_{\beta}^{'(1)}(r_1)
=
{\cal U}_1 A_{1\beta}\cos[3(k_0 r_1+\theta'_1)]
+E_{1\beta}\cos(k_0 r_1 +\theta'_1)
+F_{1\beta}\sin(k_0 r_1 +\theta'_1)
,
\label{ubeta1}
\end{eqnarray}
where the coefficients $E_{1\beta}$ satisfy
$\sum_{\beta=2}^{d}A_{1\beta}E_{1\beta}={\cal U}_3$.
Substituting the expressions (\ref{sold3}), (\ref{rtemp1}), (\ref{rdens1})
and (\ref{ubeta1}) into Eq. (\ref{secondtemp}),
and using the condition (\ref{c3}),
we eventually get
\begin{eqnarray}
\tilde{T}^{'(2)}(r_1)
&=&
{\cal T}_1 \cos[4(k_0 r_1+\theta'_1)]
+{\cal T}_2\cos[2(k_0 r_1+\theta'_1)]
\nonumber\\
&+&
{\cal T}_3\sin[2(k_0 r_1+\theta'_1)]
+{\cal T}_4
,
\end{eqnarray}
where
\begin{eqnarray}
{\cal T}_1
&=&
\frac{\hat{T}}{32}\left[
\left(
3
-8\left(\frac{\partial \kappa^{*}}{\partial\tilde{n}}\right)_0
-5\left(\frac{\partial \eta^{*}}{\partial\tilde{n}}\right)_0
\right)\hat{n}
-\frac{13}{2}\hat{T}
\right]
,
\\
{\cal T}_2
&=&
\frac{\hat{T}}{8}\biggl\{
\left(
5
+5\left(\frac{\partial \eta^{*}}{\partial\tilde{n}}\right)_0
+8\left(\frac{\partial \omega^{*}}{\partial\tilde{n}}\right)_0
\right)\hat{n}
+\frac{29}{2}\hat{T}
\nonumber\\
&+&
4d A_0^{-2}\left[\left(
1
+\left(\frac{\partial \eta^{*}}{\partial\tilde{n}}\right)_0
+2\left(\frac{\partial \omega^{*}}{\partial\tilde{n}}\right)_0
\right)\hat{n}
+\frac{3}{2}\hat{T}
\right]
\biggr\}
,
\\
{\cal T}_3
&=&
2\hat{T}A_0^{-2}\sum_{\beta=2}^{d}A_{1\beta}F_{1\beta}
,
\\
{\cal T}_4
&=&
-\frac{\hat{T}\hat{n}}{2}
.
\end{eqnarray}
Here ${\cal T}_4$ has been obtained from Eq. (\ref{c3}).
It should  be noted that ${\cal T}_1$, ${\cal T}_2$ and ${\cal T}_4$
are proportional to the square of the small parameter
$(c_v/\bar{c}_p)(P_r/d)$;
${\cal T}_3$ contains the indefinite constants $F_{1\beta}$, which
are expected to be fixed in the higher order approximations.
Generalization of this result to the case of an arbitrary
$\tilde{\bf u}^{(0)}$ given by Eq. (\ref{sold3}) is
straightforward.

In order to make analytical calculation executable in the even
higher order approximations,
it would be necessary to introduce a physically sensible assumption that
the pressure is homogeneous in the asymptotic state, 
which is in accordance with
the result (\ref{dens1}). 
Consequences of this assumption, 
however, shall not be pursued in this paper.


\begin{thebibliography}{99}

\bibitem{campbell} C.S. Campbell, Annu. Rev. Fulid Mech.
{\bf 22}, 57 (1990).

\bibitem{jaeger-nagel} H.M. Jaeger and S.R. Nagel,
Science {\bf 255}, 1523 (1992).

\bibitem{goldhirsch} I. Goldhirsch and G. Zanetti, Phys. Rev. Lett.
{\bf 70}, 1619 (1993).

\bibitem{goldhirsch2}
I. Goldhirsch, M-L. Tan and G. Zanetti, J.
Scient. Comp. {\bf 8}, 1 (1993).

\bibitem{mcnamara} S. McNamara, Phys. Fluids A {\bf 5},
3056 (1993).

\bibitem{deltour-barrat} P.Deltour and J.-L. Barrat,
J. Phys. I France {\bf 7}, 137 (1997).

\bibitem{brey1}
J.J. Brey, F. Moreno, and J.W. Dufty, Phys. Rev. E {\bf 54},
445 (1996).

\bibitem{vn-er-pre}
T.P.C. van Noije and M.H. Ernst,
Phys. Rev. E {\bf 61}, 1765 (2000).

\bibitem{brey} J.J. Brey, M.J. Ruiz-Montero, and D. Cubero,
Phys. Rev. E {\bf 60}, 3150 (1999).

\bibitem{ben-naim} E.~Ben-Naim, S.Y.~Chen, G.D.~Doolen, and S.~Redner,
Phys.  Rev. Lett. {\bf 83}, 4069 (1999).

\bibitem{soto} R. Soto, M. Mareschal, and M. Malek Mansour,
Phys. Rev. E {\bf 62}, 3836 (2000).

\bibitem{santafe} J. Wakou, R. Brito, and M.H. Ernst,
J. Stat. Phys. {\bf 107}, 3 (2002).

\bibitem{brito-ernst} R. Brito and M.H. Ernst, Europhys. Lett. {\bf 43}, 497 (1998).

\bibitem{brito-ernst2}
R. Brito and M.H. Ernst, Int. J. Mod. Phys. {\bf C 9}, 1339 (1998).

\bibitem{chen}
S. Chen, Y. Deng, X. Nie, and Y. Tu,
Phys. Lett. A {\bf 269}, 218 (2000).

\bibitem{obne} J.A.G. Orza, R. Brito, T.P.C. van Noije, and M.H. Ernst,
Int. J. Mod. Phys. C {\bf 8}, 953 (1997).

\bibitem{mcnamara-young} S. McNamara and W.R. Young,
Phys. Rev. E {\bf 53} 5089 (1996).

\bibitem{b-o} R. Brito and J.A.G. Orza, private communication.

\bibitem{tri-bar} E. Trizac and A. Barrat, Eur. Phys. J. E {\bf 3}, 291 (2000).

\bibitem{busse} F.H. Busse, J. Fluid Mech. {\bf 52}, 97 (1972).

\bibitem{jenkins-richman} J.T. Jenkins and M.W. Richman,
Phys. Fluids {\bf 28}, 3485 (1985).

\bibitem{jenkins-richman2}
J.T. Jenkins and S.B. Savage, J. Fluid Mech. {\bf 130}, 187 (1983).

\bibitem{goldstein-sapiro}
A. Goldshtein and M. Shapiro, J. Fluid Mech. {\bf 282}, 75 (1995).

\bibitem{brey-dufty-kim-santos}
J.J. Brey, J.W. Dufty, C.S. Kim, and A. Santos,
Phys. Rev. E {\bf 58}, 4638 (1998).

\bibitem{garso-dufty}
V. Garz\'o and J.W. Dufty, Phys. Rev. E {\bf 59}, 5895 (1999).

\bibitem{panton} R.L. Panton, {\it Incompressible Flow},
(Jhon Wiley \& Sons, Inc., New York, 1996).

\bibitem{haff} P.K. Haff, J. Fluid Mech. {\bf 134}, 401 (1983).

\bibitem{pa-tr-vn-er}
I. Pagonabarraga, E. Trizac, T.P.C. van Noije and M.H. Ernst,
cond-mat/0107570.

\bibitem{car-sta}
N.F. Carnahan and K.E. Starling, J. Chem. Phys. {\bf 51}, 635 (1969).

\bibitem{ver-lev}
L. Verlet and D. Levesque, Mol. Phys. {\bf 46}, 969 (1982).

\bibitem{den-krz} See, for example, P. Dennery and A. Krzywicki,
{\it Mathematics for Physicists}, (Harper \& Row, New York, 1967)
\end{thebibliography}
\end{document}